\documentclass[aps,pra,twocolumn,superscriptaddress,amsmath]{revtex4-1}

\pdfoutput = 1

\usepackage[pdftex]{graphicx,color}
\usepackage{soul}
\usepackage{amsfonts}
\usepackage{amssymb}
\usepackage{amsmath}
\usepackage{amsthm}
\usepackage{ulem}
\usepackage{enumerate}

\usepackage{subfigure}
\usepackage{rotate}
\usepackage{bm}
\usepackage{dcolumn}

\usepackage[pdftex]{hyperref}

\let\oldmarginpar\marginpar
\renewcommand\marginpar[1]{\-\oldmarginpar[\raggedleft\tiny #1]
{\raggedright\tiny #1}}

\newcommand{\avg}[1]{\left< #1 \right>}

\DeclareMathOperator{\sgn}{sgn}

\graphicspath{{figs/}}

\begin{document}

\title{The sign phase transition in the problem of interfering directed paths}

\author{C. L. Baldwin}
\affiliation{Department of Physics, Boston University, Boston, MA 02215, USA}
\affiliation{Department of Physics, University of Washington, Seattle, WA 98195, USA}

\author{C. R. Laumann}
\affiliation{Department of Physics, Boston University, Boston, MA 02215, USA}

\author{B. Spivak}
\affiliation{Department of Physics, University of Washington, Seattle, WA 98195, USA}

\date{\today}

\begin{abstract}

We investigate the statistical properties of  interfering directed paths in disordered media. 
At long distance, the average sign of the sum over paths may tend to zero (sign-disordered) or remain finite (sign-ordered) depending on dimensionality and the concentration of negative scattering sites $x$.
We show that in two dimensions the sign-ordered phase is unstable even for arbitrarily small $x$ by identifying rare destabilizing events.
In three dimensions, we present strong evidence that there is a sign phase transition at a finite $x_c > 0$.
These results have consequences for several different physical systems.
In 2D insulators at low temperature, the variable range hopping magnetoresistance is always negative, while in 3D, it changes sign at the point of the sign phase transition. 
We also show that in the sign-disordered regime a small magnetic field may enhance superconductivity in a random system of D-wave superconducting grains embedded into a metallic matrix.  
Finally, the existence of the sign phase transition in 3D implies new features in the spin glass phase diagram at high temperature. 
\end{abstract}

\maketitle

\section{Introduction} 
\label{sec:introduction}

\begin{figure}
\begin{center}
\includegraphics[width=1.0\columnwidth]{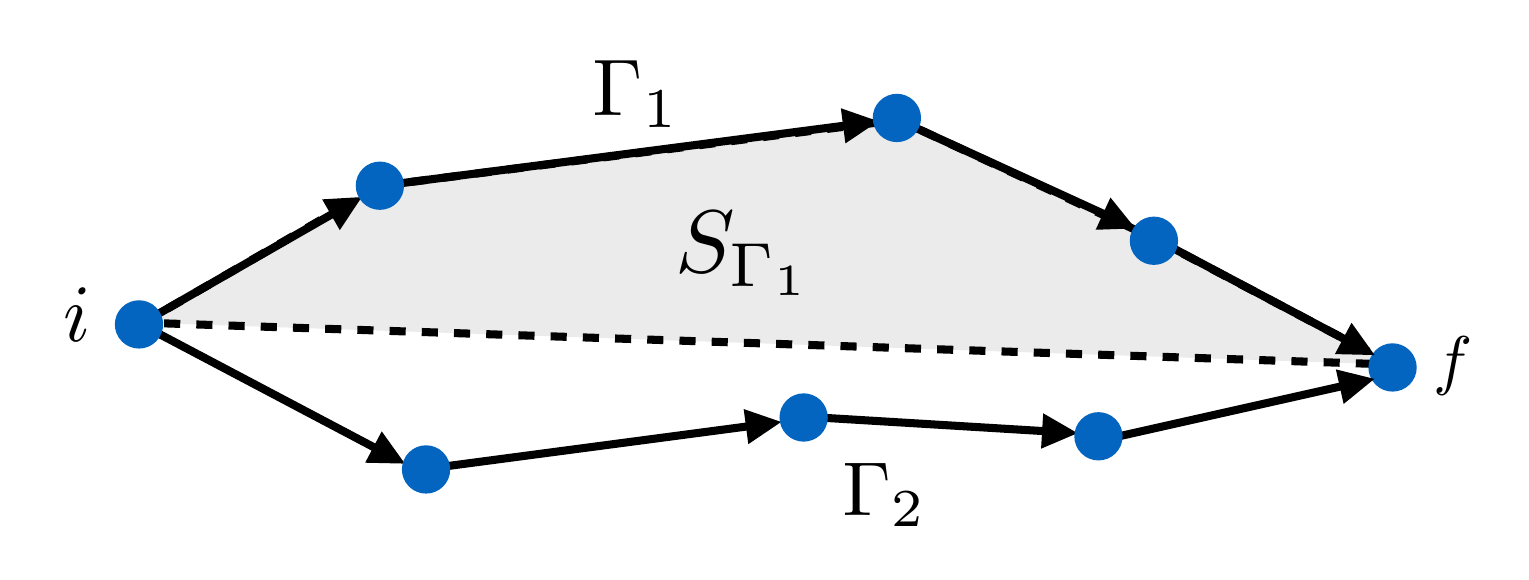}
\includegraphics[width=1.0\columnwidth]{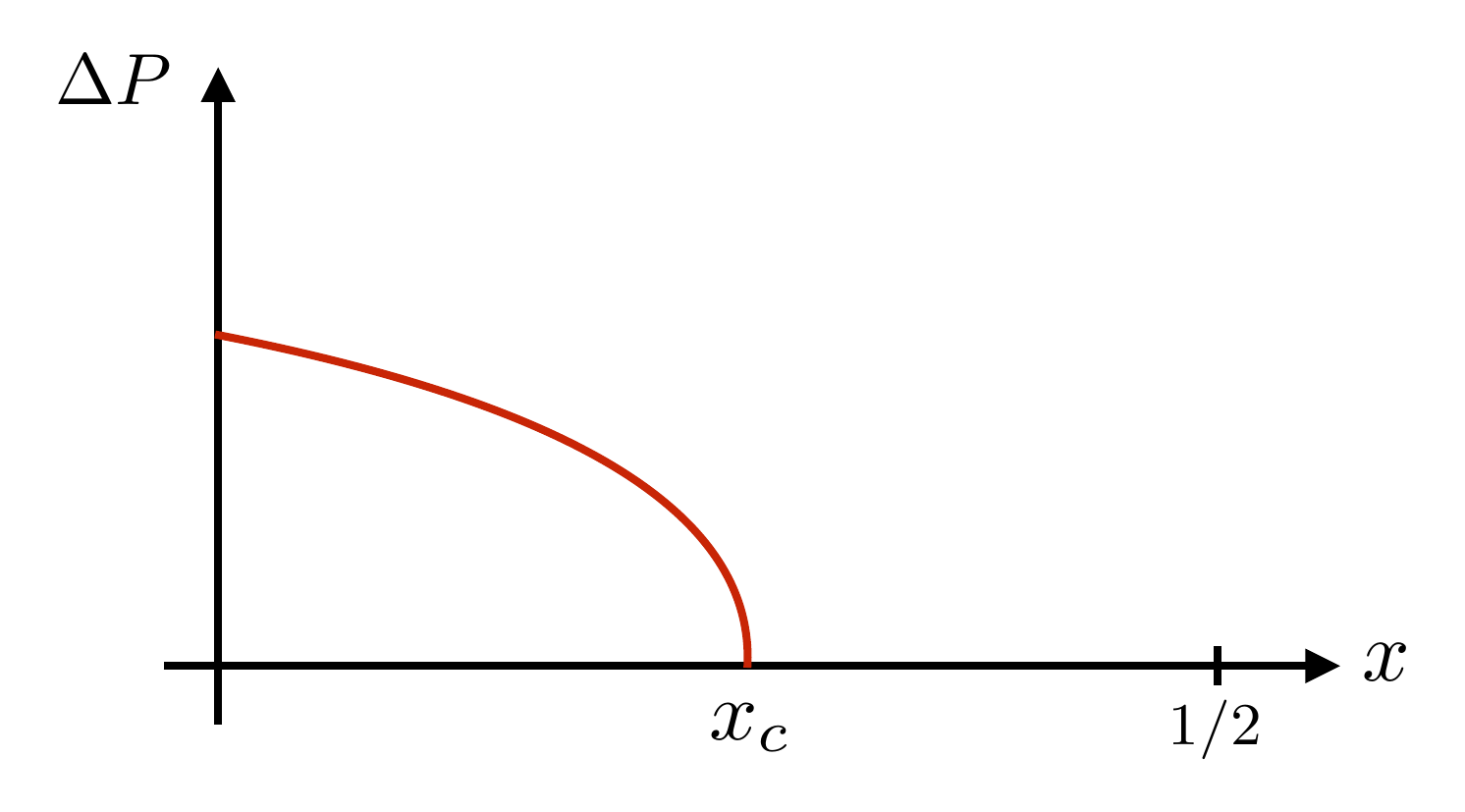}
\caption{
(top) Schematic of interfering directed paths in a random medium. $\Gamma_1$ and $\Gamma_2$ are two different paths from site $i$ to $f$. 
(bottom) Sketch of the order parameter for the sign phase transition.}
\label{fig:interfering_paths}
\end{center}
\end{figure}

In this article, we investigate the properties of interfering directed paths in random media. 
An example is shown schematically in Fig.~\ref{fig:interfering_paths}, where solid lines correspond to directed ``tunneling'' paths and blue dots represent scattering sites. We study the statistics of the sum
\begin{equation} 
	\label{eq:path_sum}
	A = \sum_{\Gamma} A_{\Gamma} , \quad A_{\Gamma} = \prod_{j \in \Gamma} \alpha_j ,
\end{equation}
where $A_{\Gamma}$ is the tunneling amplitude along path $\Gamma$, given as a product of scattering amplitudes $\alpha_j$. 
If the amplitudes $\alpha_{j}$ have random signs,  then an important property of the sum $A$ is the degree of predictability of its sign at large distances.  
This can be characterized by the difference in probabilities for $A$ to be positive and negative, respectively,
\begin{equation} \label{eq:Delta_P_definition}
\Delta P_{r\rightarrow \infty}\equiv \Delta P \equiv \textrm{Pr} [A > 0] - \textrm{Pr} [A < 0] .
\end{equation}
It was suggested in \cite{Nguyen1985Tunnel,Shklovskii1991Scattering} that the distribution function of $A$ exhibits a ``sign phase transition'' at a critical concentration of negative scattering sites $x_{c}$.
For example, if 
\begin{align}
	\label{eq:alphaDistribution}
	\alpha_{j} &= \left\{ \begin{array}{ll}
	1  & \textrm{with probability } 1-x \\
	-M & \textrm{with probability } x
	\end{array} \right. 
\end{align}
then
\begin{align}
\begin{array}{lll}
	\Delta P > 0 &\textrm{ for } x< x_c, \\
	\Delta P = 0 &\textrm{ for } x> x_c.
\end{array}
\end{align}
The quantity $\Delta P(x)$ serves as an order parameter for the sign phase transition.
Such a transition is shown qualitatively in the bottom of Fig.~\ref{fig:interfering_paths}.

It was argued in \cite{Shklovskii1991Scattering} that the upper critical dimension for the sign phase transition is four.
The lower critical dimension has been widely debated, and in particular, whether the sign-ordered phase exists in 2D has been a controversial subject for a long time \cite{Shapir1987Absence,Medina1989Interference,Medina1992Quantum,Roux1994Interference,Nguyen1996Crossover,Shklovskii1991Scattering,Spivak1996Sign,Aponte1998Directed,Kim2011Interfering,Ioffe2013Giant}.

Here, we show conclusively that the sign-ordered phase does not exist in 2D ($\Delta P = 0$ for any $x > 0$), and present strong numerical evidence that it does exist in 3D.  
The former result is consistent with some of the previous studies (see in particular~\cite{Kim2011Interfering}). 
We explain the instability of the sign-ordered phase at small values of $x$ by identifying the rare fluctuations which destabilize the sign order. 
These lead to an anomalously large correlation length which scales stretched-exponentially with $x$ and explains the apparent sign order observed in previous numerical studies \cite{Nguyen1985Tunnel,Medina1992Quantum,Spivak1996Sign,Ioffe2013Giant}.

The (non-)existence of the sign-ordered phase and associated transition has immediate consequences for the following physical systems:
\begin{enumerate}[a)]

\item 

The quantity $A$ in Eq.~\eqref{eq:path_sum} can play the role of the electron tunneling amplitude in a disordered medium, where it arises as a sum of partial amplitudes corresponding to different tunneling paths  
\cite{Shklovskii1985Effect,Nguyen1985Tunnel,Sivan1988Orbital,Medina1990Magnetic,Shklovskii1991Scattering,Zhao1991Negative,Gangopadhyay2013Magnetoresistance}.
It was argued in Refs.~\cite{Nguyen1985Tunnel,Shklovskii1991Scattering,Zhao1991Negative} that the sign of the magnetoresistance in the hopping conductivity regime depends on whether the system is in the sign-ordered or -disordered phase. 
 
\item 
In the Edwards-Anderson spin glass, the spin correlation function at high temperature is governed by a sum analogous to $A$, where the $\alpha_j$ correspond to bond disorder. 
Thus, the presence of a sign-ordered phase in 3D implies that a transition takes place in the sign of the correlation functions at high temperature.

\item 
At high temperature, the correlation function $\chi=\avg{\exp{(i (\phi_i - \phi_j))}}$ in a system  of randomly oriented and randomly shaped grains of D-wave superconductor embedded into a metallic matrix can be reduced to Eq.~\eqref{eq:path_sum}.  
Here,  $\phi_i$ is the phase of the order parameter on grain $i$.  
In analogy with the magnetoresistance in the hopping conductivity regime, the magnetic field suppresses superconductivity in the sign-ordered phase and enhances it in the sign-disordered phase. 
\end{enumerate}
We will return to these applications in more detail in Sec.~\ref{sec:applications}.

In the following, we first review the essential physical picture of the sign-ordered phase in Sec.~\ref{subsec:SPT_MF}. 
We then develop a more detailed picture of the fluctuations in 2D which lead to the instability of sign order in Sec.~\ref{subsec:SPT_2D}, and confirm these with numerical simulations. 
In Sec.~\ref{subsec:SPT_3D} we turn to the 3D problem and present evidence that sign order is stable at small $x$.
Finally, we present applications and discussion in Secs.~\ref{sec:applications} and~\ref{sec:conclusion}.

\section{The sign phase transition} 
\label{sec:SPT}

\subsection{Mean-field description and generalities} 
\label{subsec:SPT_MF}

\begin{figure}
\begin{center}
\includegraphics[width=1.0\columnwidth]{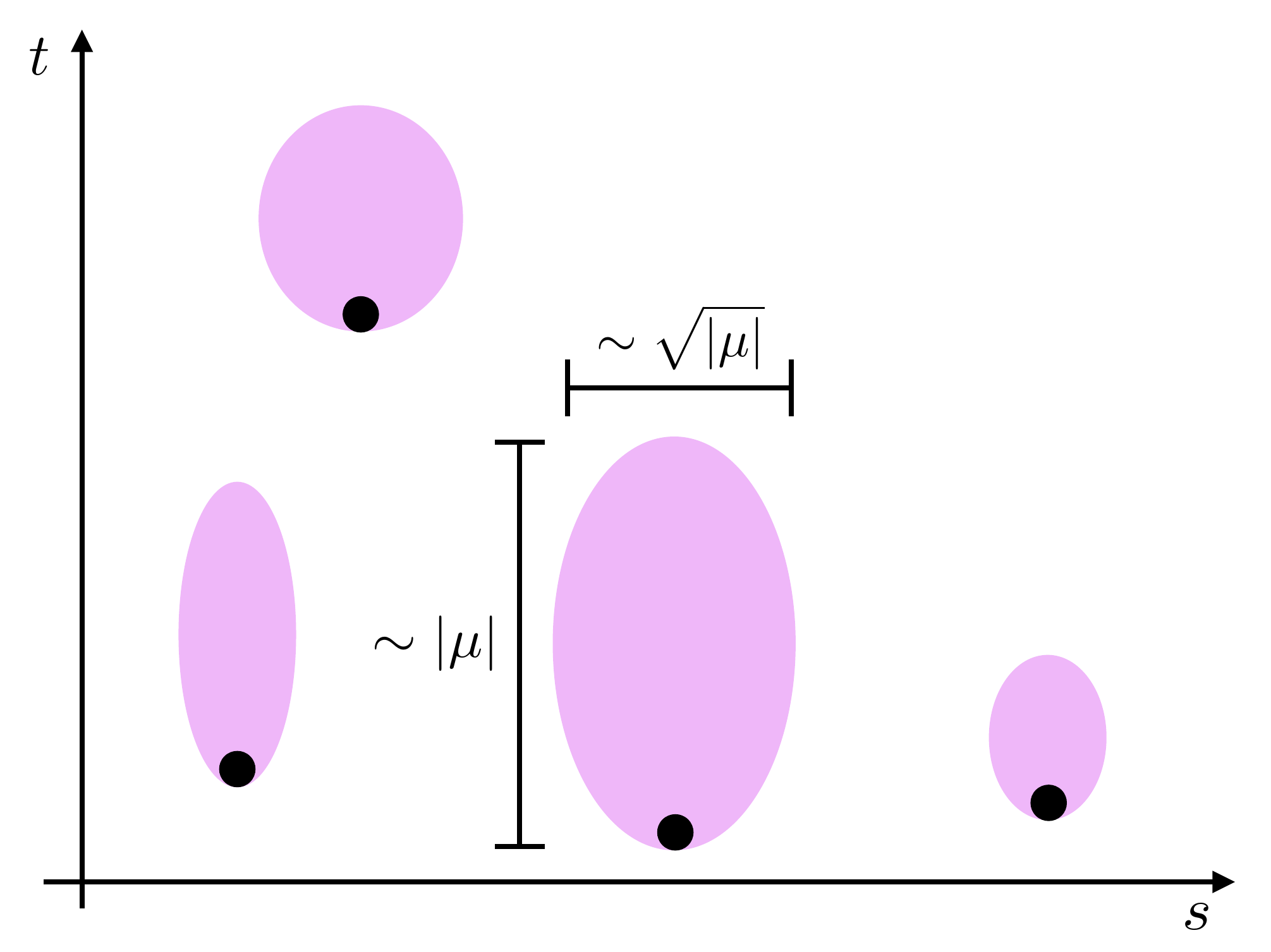}
\caption{
Regions in which the path sum is negative (purple), created by isolated negative scatterers (black dots). 
Although each region has a random size, the typical scales are as shown. The paths are directed along $t$. 
The scattering length $\mu$ characterizes the strength of the scatterer, see~\cite{Ioffe2013Giant} for details.}
\label{fig:cigar_shaped_region}
\end{center}
\end{figure}
 
The essential picture of the sign-ordered phase is illustrated in the ``space-time'' diagram of Fig.~\ref{fig:cigar_shaped_region}.
Here, the ``time'' coordinate $t$ corresponds to the direction of propagation of the directed paths and the ``spatial'' coordinates $s$ to the $d-1$ transverse directions.
Negative-amplitude scatterers produce cigar-shaped negative domains in the sign 
\begin{align}
	\sigma(s,t) \equiv \sgn[A(s,t)]
\end{align}
of the amplitude field. 

For an isolated negative scatterer in an otherwise uniform lattice of positive scatterers, the sign at $(s,t)$ is determined by the interference between those paths which go through the negative scatterer and those which miss it.
If the scattering amplitude is sufficiently large, the path sum may be estimated in the diffusive limit,  
\begin{equation}
	A(s,t) \propto 1 - \left( \frac{|\mu|}{t} \right)^\frac{d-1}{2} e^{-\frac{s^2}{4 D t}} 
\end{equation}  
where the scattering length $\mu$ characterizes the strength of the negative scatterer, $D$ is a microscopic length, and we have suppressed an $O(1)$ constant.
We find that the negative domain $A < 0$ has extent  $\tau \sim |\mu|$, width $w \sim \sqrt{|\mu|}$, and volume $\mathfrak{v} \sim |\mu|^{\frac{d+1}{2}}$. 
At sufficiently small concentration of scatterers $x$, the negative domains remain far apart and do not interfere.  
The sign field $\sigma(s, t)$ only disorders if the domains percolate, i.e., when
\begin{equation}
x \mathfrak{v} = x |\mu|^{\frac{d+1}{2}}  \gtrsim 1. 
\end{equation}
Thus this picture predicts a finite $x_c$ for sign order in any dimension $d>1$ \footnote{In $d=1$, randomly placed single scatterers clearly disorder the sign field.}.

\begin{figure*}[t]
\begin{center}
\includegraphics[width=1.0\textwidth]{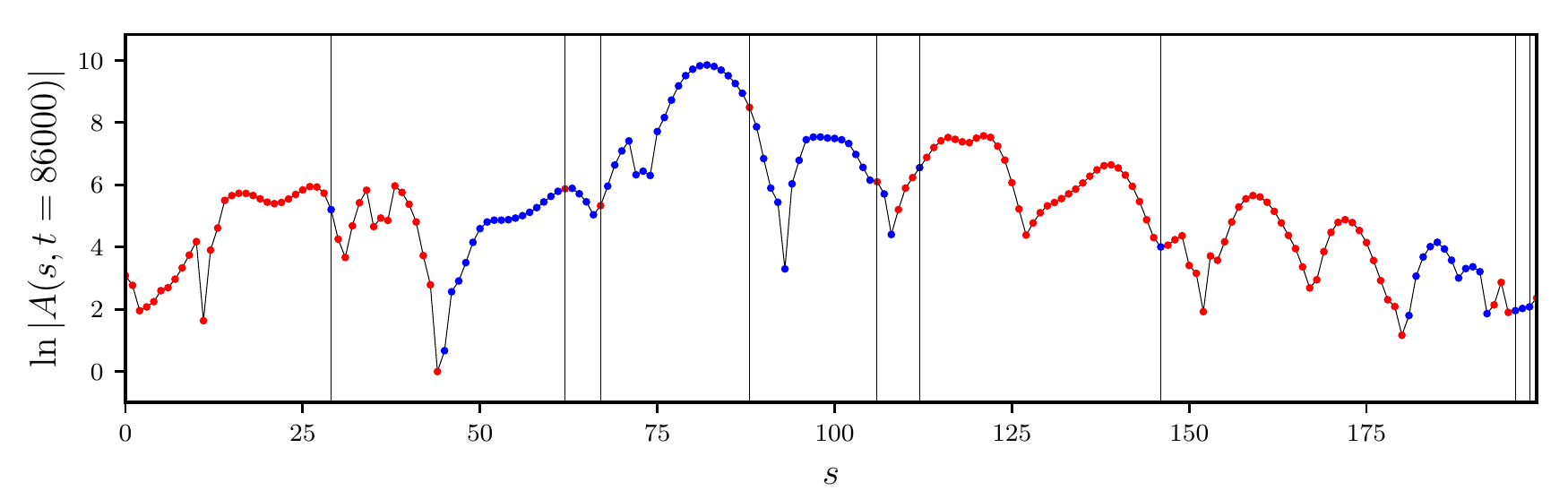}
\caption{
A typical snapshot of the (log-)amplitude field $\ln{|A(s, t)|}$, as a function of $s$ for fixed $t$ ($L=200$, $x=0.04$). 
The $y$-axis values are shifted so that the minimum is at 0. 
The time $t = 86000$ is close to the disordering timescale $t^*$ for this $x$. 
Red (blue) points correspond to positive (negative) $\sigma(s,t) \equiv \sgn[A(s,t)]$. 
The vertical black lines indicate the negative scatterers at the current $t$, i.e., where $\alpha_{(s, t)} = -1$. 
All other sites have $\alpha_{(s, t)} = 1$.}
\label{fig:amplitude_field}
\end{center}
\end{figure*}

This argument neglects fluctuations in the size of the isolated negative domains. 
Should the distribution of domains have a sufficiently long tail, then sign order becomes unstable even at very small $x$, as argued by \cite{Kim2011Interfering}.
Suppose that the distribution of domain lengths $\tau$ has a power-law tail, $p_s(\tau) \sim \tau^{-\eta}$, and that the typical transverse width of such domains is $w(\tau) \sim \tau^{\gamma}$.
We refer to $\eta$ as the ``survival'' exponent and $\gamma$ as the ``growth'' exponent.
The fraction of the transverse volume occupied by negative domains at time $t$ is
\begin{align}
\label{eq:spacefrac}
	x\int^t dt'  p_s(t-t') w(t-t')^{d-1}.
\end{align}
This fraction converges as $t \rightarrow \infty$ provided
\begin{equation} \label{eq:exponent_inequality}
1 - \eta + (d-1) \gamma < 0.
\end{equation}
Inequality \eqref{eq:exponent_inequality} is a necessary condition for the stability of the sign-ordered phase with respect to these fluctuations. 

It is instructive to interpret the sign $\sigma(s,t)$ as an Ising field in $d-1$ spatial dimensions $s$ and temporal dimension $t$, which evolves in the presence of ``noise'' given by the scattering disorder.
In this language, the sign order parameter is simply the magnetization as $t \rightarrow \infty$, 
\begin{equation} \label{eq:sign_order_parameter_magnetization}
\begin{aligned}
\mathbb{E}[\sigma(s, t)] =& \, \textrm{Pr}[A(s, t) > 0] - \textrm{Pr}[A(s, t) < 0] \\
=& \, \Delta P(s, t)  .
\end{aligned}
\end{equation}
Here $\mathbb{E} [ \cdot ]$ denotes averaging with respect to the random distribution of scatterers (i.e., the noise).
Since the lower critical (spatial) dimension of the equilibrium Ising model is $1$, we might expect the sign-ordered phase to be unstable to fluctuations for $d \le 2$ and stable for $d \ge 3$.
We will argue below that this is indeed the case, despite that the noise from scattering does not obey detailed balance with respect to an equilibrium Ising model.

\subsection{Absence of the sign-ordered phase in 2D} 
\label{subsec:SPT_2D}

\subsubsection{Survival and growth exponents} \label{sssec:survival_growth_exponents}

As a warm-up, consider the 1D equilibrium Ising model with Glauber dynamics at low temperature.
Domain walls undergo random walks and annihilate when they meet.
When a single spin is flipped in a uniform background, the resulting domain has probability $p_s(\tau) \sim \tau^{-1/2}$ of surviving until time $\tau$. 
Over that time, the walls typically walk $w \sim \tau^{1/2}$.
Thus, $\eta = \gamma = \frac{1}{2}$ and inequality~\eqref{eq:exponent_inequality} with $d-1 = 1$ is violated.
The magnetization is unstable, as expected for a finite-temperature 1D model.

In the directed path problem, there is no equilibrium for the sign field $\sigma$. The stochastic ``dynamics'' nevertheless induce survival and growth exponents.
We follow \cite{Kim2011Interfering} and consider an isolated negative scatterer embedded in a dense background of \textit{disordered} positive scatterers.
The path sum in the positive-scattering background reduces to the well-known directed polymer problem \cite{Fisher1991Directed,HalpinHealy1995Kinetic}. 
In the extreme disordered limit, the polymer ``pins'' so that one path $\Gamma_0$ dominates the sum:
\begin{align}
    A = A_{\Gamma_0} + \cdots .
\end{align}
Accordingly, the sign $\sigma(s,t)$ is only negative if $\Gamma_0$ happens to go through the lone negative scatterer. 
It is known that the directed polymer wanders over a distance $w(\tau) \sim \tau^\xi$ with wandering exponent $\xi = \frac{2}{3}$.
Thus, we identify $\eta = (d-1)\xi$ and $\gamma = \xi$.
Inequality~\eqref{eq:exponent_inequality} is violated, which again implies the instability of the sign-ordered phase even at arbitrarily small $x$.

Since the fraction of space occupied by negative domains at time $t$ is $xt$ (see Eq.~\eqref{eq:spacefrac}), we also obtain a simple estimate for the disordering time: 
\begin{equation}
    \label{eq:davidtstar}
    t^*(x) \sim x^{-1}.
\end{equation}

This argument clearly applies to the large-disorder limit where the path sum is dominated by a single path $\Gamma_0$. 
At weaker disorder in the ``pinned'' phase, the wandering exponent $\xi$ governing the directed polymer is unchanged, yet subdominant paths now contribute to the path sum and interference effects may become nontrivial.
Numerical investigations in Ref.~\cite{Kim2011Interfering} confirmed that the survival and growth exponents for the domains produced by isolated negative scatterers are nevertheless unchanged when the background disorder is of intermediate strength in 2D.

\subsubsection{Negative scatterers and the role of interference} \label{sssec:negative_scatterers}

The above analysis relies on positive background disorder to produce the destabilizing fluctuations.
It leaves open the possibility of 2D sign order when the disorder arises only from negative scatterers.
Here we close the door by considering this regime in the limit where the typical negative domain is microscopic ($\mu \lesssim D$) and the concentration of negative scatterers $x \rightarrow 0$.
We find that sign order is nonetheless destroyed by rare events.

First consider no disorder ($x=0$). The sum Eq.~\eqref{eq:path_sum} describes  diffusion of paths, so that in the continuum limit, 
\begin{align}
	\partial_t A = D\nabla_s^2 A 
\end{align}
where we have rescaled $A$ exponentially with $t$ in order to remove an overall $s$-independent factor. 
Suppose the amplitude at $t=0$ is roughly uniform over a region of width $l$. If a negative scatterer flips the sign of $A$ in a subregion of width $w \ll l$, the negative domain becomes positive after a time $\tau \sim w^2/D$.
Thus, isolated negative scatterers do not produce asymptotically long-lived negative domains.

For small but finite concentration $x$, a large length scale $l(x)$ emerges.
Fig.~\ref{fig:amplitude_field} shows a typical realization of the amplitude $A(s, t)$ at late time $t$ (see Sec.~\ref{sssec:numerical_validation} for numerical details). 
The log-amplitude field forms smooth ``hills'' separated by sharp minima.
The length scale $l(x)$ is the typical distance between minima, i.e., the typical width of a hill.
As in the isolated case, scattering events which produce negative domains of width $w \ll l(x)$ remain short-lived ($\tau \sim w^2$).
However, if a negative scattering event produces a domain covering more than half of the weight in the hill ($w \sim l(x)/2$), then it cannot disappear due to diffusion of amplitude within the hill. 
Such domains are locally stable and their lifetimes are governed by competition with neighboring hills over much longer timescales.
Thus, $l(x)$ separates short-lived and long-lived domains.


A self-consistent argument gives the scaling of $l(x)$ as $x \rightarrow 0$.
A single negative scatterer at time $t_0$, although it does not produce a lasting negative domain, creates a local minimum in $|A(s, t)|$.
The minimum becomes wider and shallower as $\Delta t \equiv t - t_0$ increases, with the width scaling as $\Delta t^{1/2}$.
After a time $\Delta t \sim l(x)^2$, the minimum merges into its neighbors and can no longer be resolved.
Thus  $l(x)^2$ is the ``lifetime'' of a local minimum.
New minima are created at a rate $x$ per unit length and time.
Thus, the typical density of minima present at any given time is $\sim xl(x)^2$, but by definition, this must equal $\frac{1}{l(x)}$.
We have that
\begin{align}
    \label{eq:length_scaling}
    l(x) &\sim x^{-\frac{1}{3}}
\end{align}
which holds for $x \ll 1$. 

Coarse-grained on the scale $l(x)$, isolated negative scatterers become effective positive-weight disorder, while the rare events which produce domains of width $l(x)$ become negative scatterers whose concentration is $x^{\alpha l(x)}$ for some $O(1)$ constant $\alpha$. 
On this scale, the effective positive scatterers are disordered, so the analysis of Sec.~\ref{sssec:survival_growth_exponents} and Ref.~\cite{Kim2011Interfering} again applies.
We recover that the sign order is unstable but with a parametrically longer timescale (cf. Eq.~\eqref{eq:davidtstar}),
\begin{align} 
    \label{eq:disordering_time}
    t^*(x) &\sim  \; x^{-\alpha l(x)} \sim x^{-\alpha x^{-1/3}},
\end{align}
for $x \to 0$.

\subsubsection{Numerical validation} 
\label{sssec:numerical_validation}

The above arguments are qualitative and require numerical validation, which we now present.

\begin{figure}[t]
\begin{center}
\includegraphics[width=1.0\columnwidth]{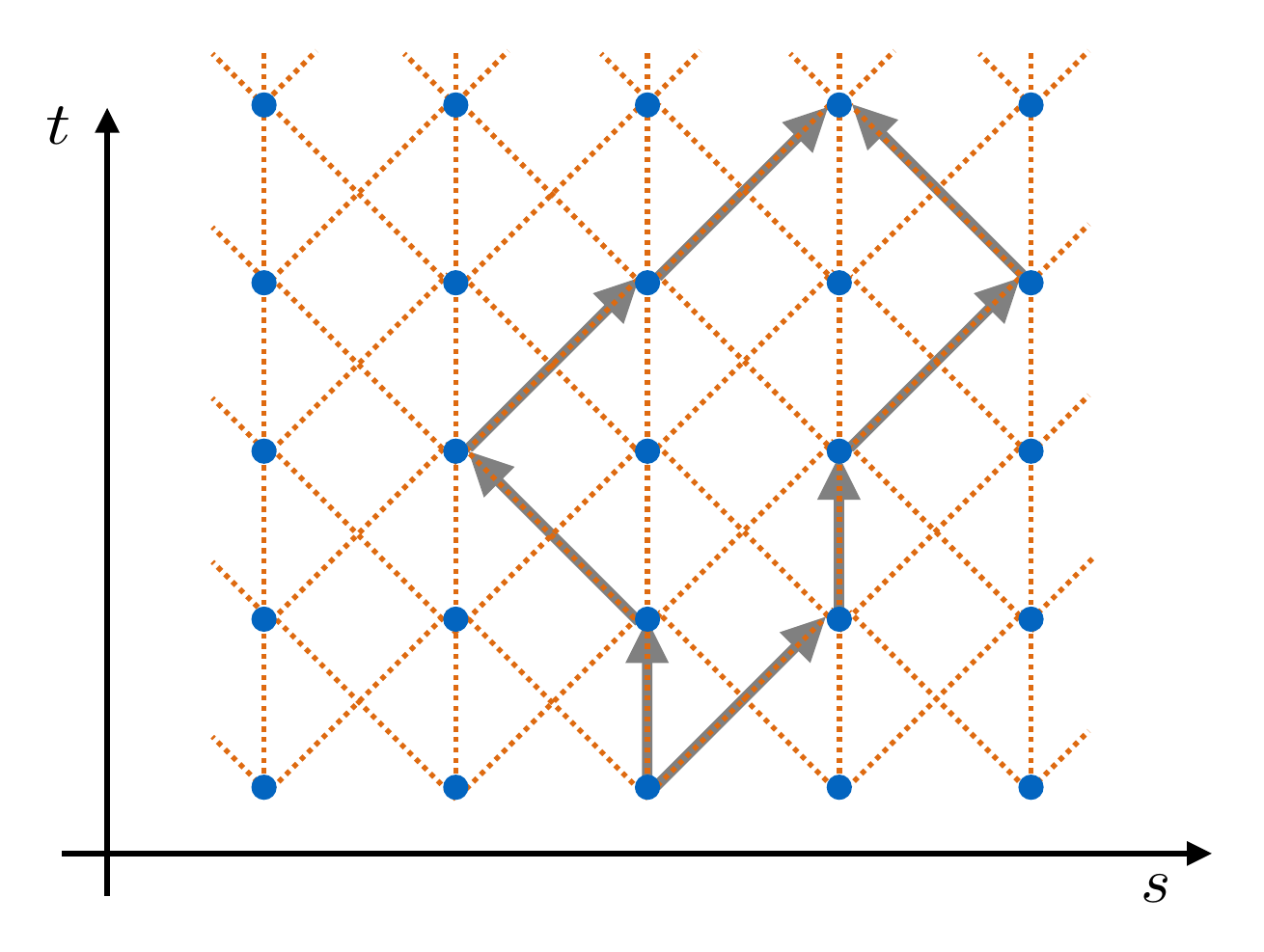}
\caption{
The lattice (blue dots) used in the 2D simulations. 
All 2D simulations begin with uniform initial conditions $A(s, 0) = 1$ and propagate forward in $t$ using Eq.~\eqref{eq:trident_recursion}. The arrows show two examples of directed paths on this lattice.}
\label{fig:lattice}
\end{center}
\end{figure}

For concreteness, we use the lattice shown in Fig.~\ref{fig:lattice}.
Each site $(s, t)$ contains a scatterer with random amplitude $\alpha_{(s, t)}$, for which we take the binary distribution given by Eq.~\ref{eq:alphaDistribution} with $M=1$.
Instead of evaluating each path amplitude $A_{\Gamma}$, we organize the sum over paths iteratively:
\begin{align} 
    \label{eq:trident_recursion}
    A(s, t+1) = \alpha_{(s, t+1)} & \left( A(s-1, t) \right. \nonumber \\
    & \qquad \left. + A(s, t) + A(s+1, t) \right) .
\end{align}
In all simulations in this section, we consider ``quenches'' from uniform initial conditions $A(s, 0) = 1$ in systems with transverse width $L$ and periodic boundary conditions.

\begin{figure}
\begin{center}
\includegraphics[width=1.0\columnwidth]{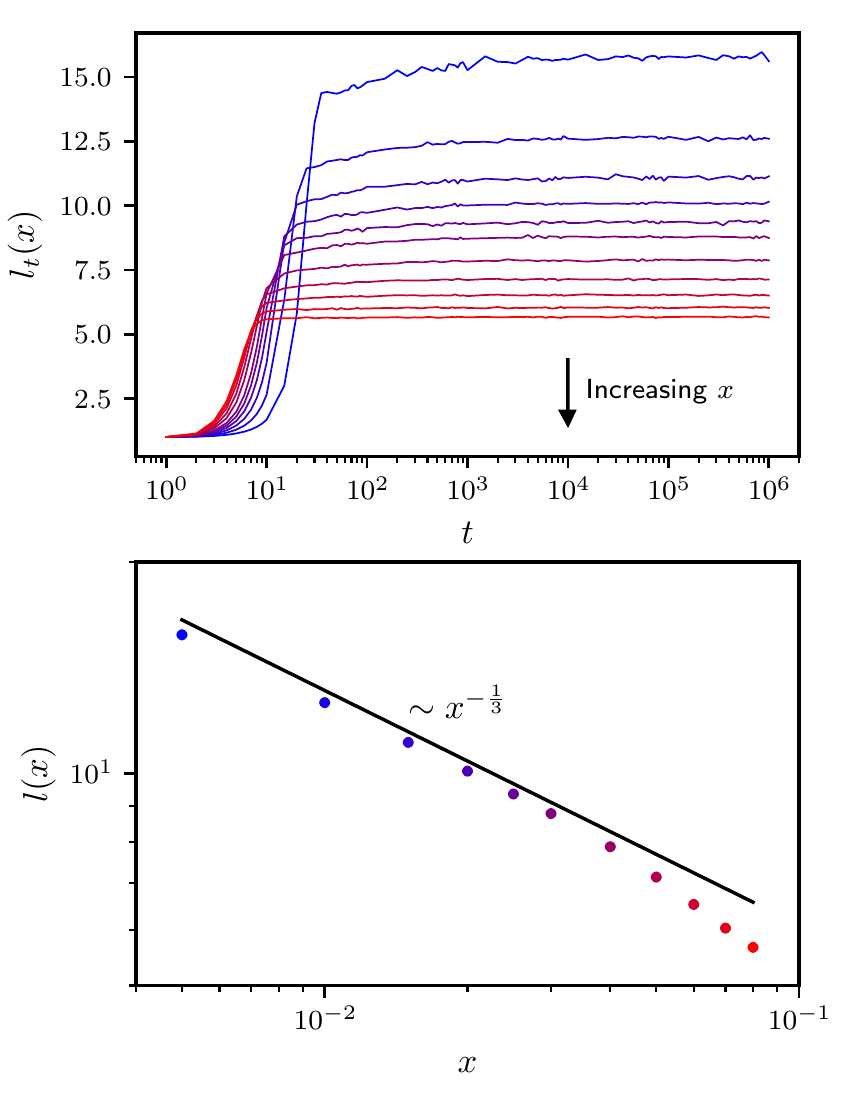}
\caption{
(top) 
Mean distance between local minima of the height field at time $t$, averaged over disorder ($L =1600$).
Each line (color) corresponds to a different density of scatterers $x$.
Errorbars (not shown for clarity) are of the same magnitude as the fluctuations within each curve. 
(bottom) 
Asymptotic distance $l(x)$ between minima, taken from the average of the late-time plateaux of $l_t(x)$. 
The color of each point indicates which curve in the top panel it corresponds to. 
Errorbars are smaller than the marker size. 
The solid line is a power-law curve, drawn to guide the eye.}
\label{fig:parabola_width}
\end{center}
\end{figure}

We first determine $l(x)$ numerically by defining, at fixed time $t$, $l_t(x)$ to be the disorder- and spatial-averaged distance between local minima of $\ln{|A(s, t)|}$. 
Fig.~\ref{fig:parabola_width} shows $l_t(x)$ as a function of $t$ for a system of size $L=1600$ (the curves are independent of $L$).
Since the curves saturate at $t$ well within the simulation time, we determine $l(x) \equiv \lim_{t \rightarrow \infty} l_t(x)$ by averaging the $l_t(x)$ over their plateaux. 
The scaling behavior of the resulting $l(x)$ with $x$, shown in Fig.~\ref{fig:parabola_width}b, confirms Eq.~\ref{eq:length_scaling}.

\begin{figure}
\begin{center}
\includegraphics[width=1.0\columnwidth]{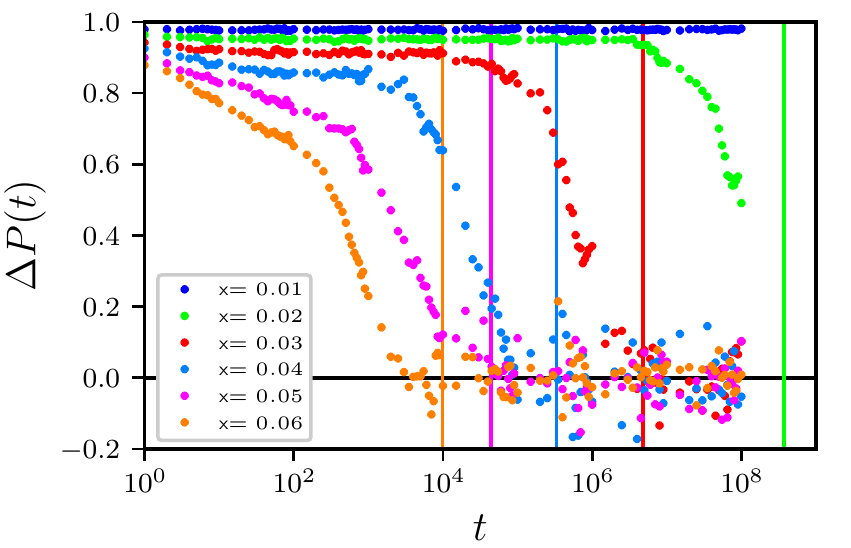}
\caption{
Decay of the sign order parameter $\Delta P(t) \equiv \mathbb{E} \left[ \frac{1}{L} \sum_s \sigma(s, t) \right]$ in 2D from uniform initial conditions (transverse size $L=100$). 
Statistical errorbars (not shown) are of the same magnitude as the fluctuations within a curve. 
The vertical lines correspond to the independently determined disordering timescales $t^*(x)$ for each density of negative scatterers $x$, see Eq.~\eqref{eq:disordering_time}. 
The predicted $t^*(x=0.01) \approx 10^{12}$ is not accessible with current computing resources.
}
\label{fig:sign_disordering}
\end{center}
\end{figure}

We have also verified Eq.~\eqref{eq:disordering_time} for the disordering time.
Fig.~\ref{fig:sign_disordering} shows
\begin{equation} \label{eq:Delta_P_def}
\Delta P(t) \equiv \mathbb{E} \left[ \frac{1}{L} \sum_s \sigma(s, t) \right]
\end{equation}
as a function of $t$, for various small $x$.
For all $x \geq 0.02$, the sign field clearly disorders at large $t$.
The vertical lines are the independent estimates $t^*(x) = x^{-\alpha l(x)}$, with $\alpha = \frac{1}{2}$ and $l(x)$ determined numerically as described above.
The agreement with the observed disordering times is excellent considering that $t^*(x)$ ranges over $\sim 5$ orders of magnitude as $x$ varies.
This also explains why past numerical work on the 2D sign phase transition was inconclusive: the simulation must run for very long times to see disordering.
Indeed, we estimate that the disordering time for $x = 0.01$ is $\sim 10^{12}$, which is longer than we can study numerically.

We note that within our analysis, $\alpha$ is the only free fitting parameter. 
$\alpha = \frac{1}{2}$ gives an excellent fit and has a simple physical rationale: only half of a hill must change sign simultaneously, for then the new domain occupies the majority of the hill and annihilates the remainder.

\subsection{The sign-ordered phase in 3D} 
\label{subsec:SPT_3D}

There are several suggestive but contradictory arguments regarding the sign-ordered phase in 3D. 
The analogy with the $(d-1)$-dimensional stochastic Ising model (see Sec.~\ref{subsec:SPT_MF}) suggests that the sign-ordered phase can exist, since Ising order is stable in two spatial dimensions.
On the other hand, disorder always drives the (positive-weight) directed polymer into its ``pinned'' phase in 3D \cite{HalpinHealy1995Kinetic}, just as in 2D. 
In the strongly pinned limit where $A$ is dominated by a single path, this would lead to sign disorder by the arguments of Sec.~\ref{sssec:survival_growth_exponents} and Ref.~\cite{Kim2011Interfering}. 
However, this does not rule out the possibility of a stable sign-ordered phase at weaker disorder.
Here, we present a numerical study in the weak disorder regime analogous to that studied in 2D above.
By several complimentary numerical simulations and finite-size scaling analyses, we  conclude that 3D sign order exists.

We calculate path sums on the cubic lattice defined by the recursion relation
\begin{align} \label{eq:3D_lattice_recursion}
A(s_1, s_2, t+1) =& \; \alpha_{(s_1, s_2, t+1)} \Big( A(s_1, s_2, t)  \nonumber \\
+ & \; A(s_1 - 1, s_2, t) + A(s_1 + 1, s_2, t) \\
+ & \; A(s_1, s_2 - 1, t) + A(s_1, s_2 + 1, t) \Big) , \nonumber
\end{align}
with periodic boundary conditions for systems of transverse size $L$ x $L$.

\begin{figure}
\begin{center}
\includegraphics[width=1.0\columnwidth]{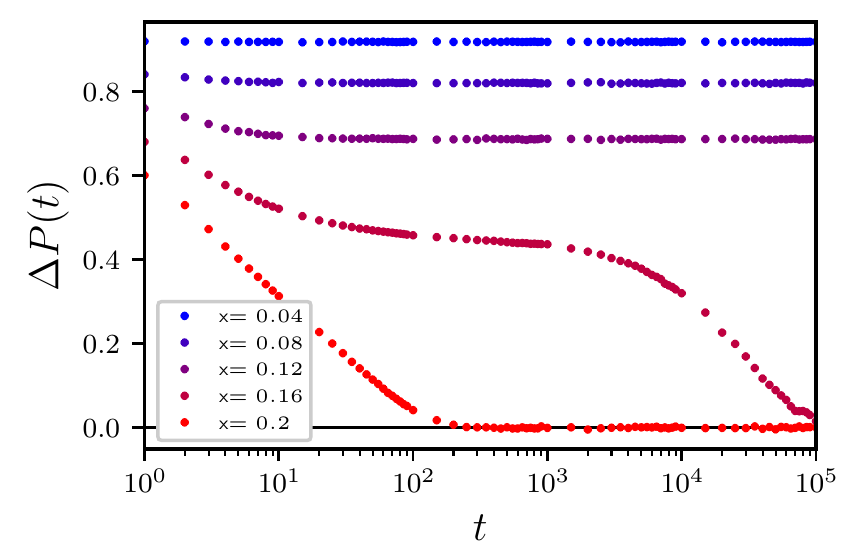}
\includegraphics[width=1.0\columnwidth]{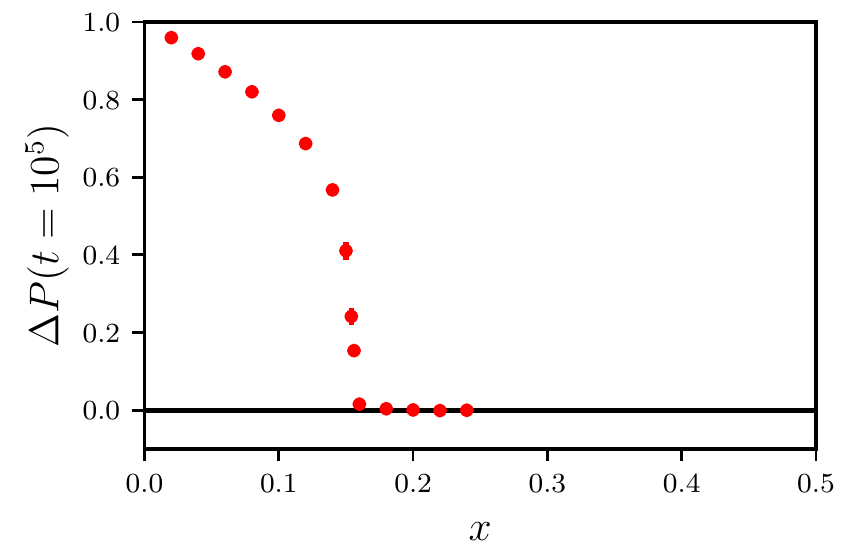}
\caption{
Decay of sign order in 3D from uniform initial conditions on the lattice of Eq.~\eqref{eq:3D_lattice_recursion}.
(top) The sign order parameter $\Delta P(t)$ as a function of $t$ for various $x$ at size $L=80$. 
Statistical errorbars are smaller than the marker size. 
(bottom) $\Delta P$ evaluated at time $t = 10^5$ as a function of $x$. The sharpness of the crossover is suggestive of a transition at $x_c \approx 0.16$.
}
\label{fig:3D_decay_order}
\end{center}
\end{figure}

Fig.~\ref{fig:3D_decay_order} shows the decay of $\Delta P(t)$ starting from uniform initial conditions. 
It suggests that the sign field becomes disordered when $x \gtrsim 0.16$ but remains ordered when $x \lesssim 0.16$. 
However, we face the same difficulty as in 2D, (cf. Fig.~\ref{fig:sign_disordering}): $\Delta P(t)$ may remain non-zero throughout the accessible simulation but disorder on longer timescales.

\begin{figure}
\begin{center}
\includegraphics[width=1.0\columnwidth]{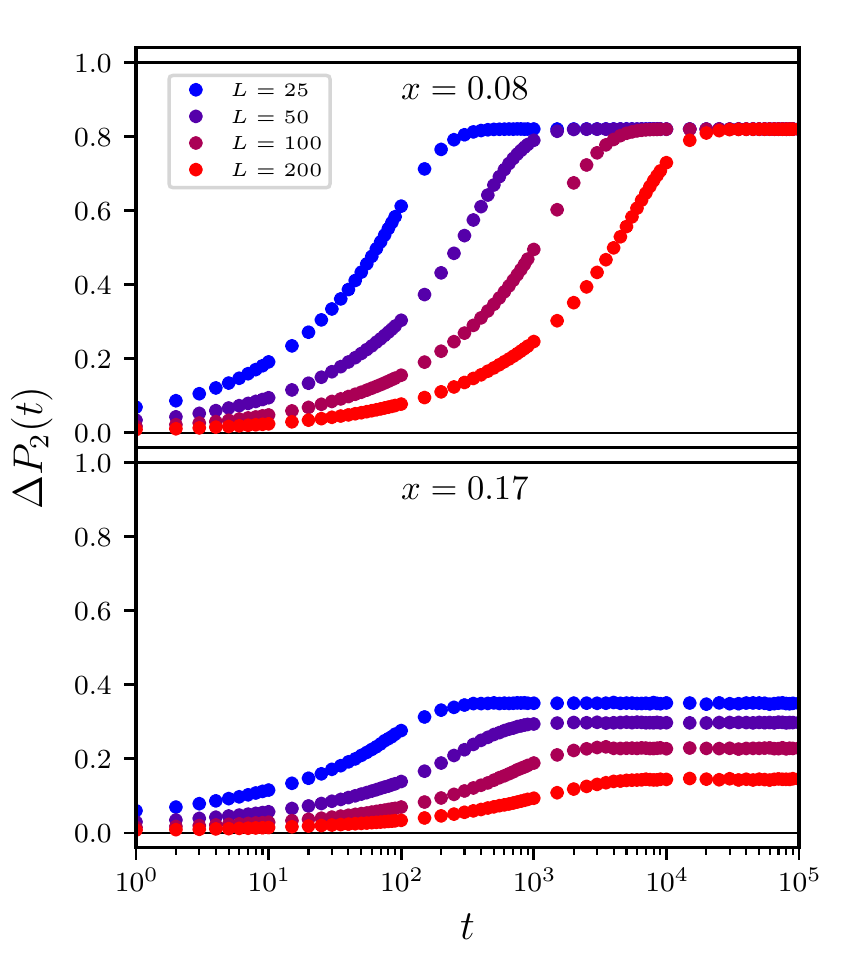}
\caption{
Spontaneous ordering of the sign field starting from disordered initial conditions, as shown by $\Delta P_2(t)$ (Eq.~\eqref{eq:alternate_order_parameter}) for various $L$. 
Statistical errorbars are smaller than the marker size.
(top) For $x=0.08$, the late-time value of $\Delta P_2(t)$ becomes independent of system size $L$, consistent with spontaneous long-range order.
(bottom) For $x=0.17$, the late-time value of $\Delta P_2(t)$ decays with increasing system size.}
\label{fig:3D_quench_order}
\end{center}
\end{figure}

To confirm that the sign-ordered phase is in fact stable at small $x$, we consider ``quench'' experiments from \textit{disordered} initial conditions: $A(s_1, s_2, 0) = \pm 1$ with equal probability.
If the sign order is stable, we expect the sign field to order spontaneously for $x < x_c$. 
This is in analogy to the 2D Ising model, which magnetizes spontaneously when quenched from high temperature to below $T_c$. 
Fig.~\ref{fig:3D_quench_order} demonstrates this ordering for two representative concentrations $x$. 
Note that because of the symmetry in the initial conditions, we consider the order parameter
\begin{equation} \label{eq:alternate_order_parameter}
\Delta P_2(t) \equiv \sqrt{\mathbb{E} \left[ \left( \frac{1}{L^2} \sum_{s_1, s_2} \sigma(s_1, s_2, t) \right) ^2 \right] }.
\end{equation}
At $x = 0.08$, well below the tentative $x_c$ identified above, $\Delta P_2(t)$ approaches a constant value independent of $L$ as $t \rightarrow \infty$. 
The timescale to reach the asymptotic value scales as $L^2$ (not shown), which is the same scaling as that of coarsening dynamics in the 2D Ising model \cite{Cugliandolo2010Topics}. 
At $x = 0.17$, in contrast, $\lim_{t \rightarrow \infty} \Delta P_2(t)$ decreases as the system size increases, consistent with lack of long-range order.

These two complementary simulations, respectively observing the decay of ordered sign fields and the spontaneous ordering of disordered ones, together suggest that the sign field remains ordered at small $x$ and only disorders at larger $x$. 
To quantitatively extract the critical $x_c$, we have carried out a crossing-point analysis of the Binder cumulant obtained from the uniform-initial-condition simulations. 
The sign Binder cumulant
\begin{equation} \label{eq:binder_cumulant}
U(t) \equiv 1 - \frac{\mathbb{E} \left[ \left( \frac{1}{L^2} \sum_{s_1, s_2} \sigma(s_1, s_2, t) \right) ^4 \right] }{\mathbb{E} \left[ \left( \frac{1}{L^2} \sum_{s_1, s_2} \sigma(s_1, s_2, t) \right) ^2 \right] ^2 }
\end{equation}
provides a dimensionless measure of the ordering transition in the sign field~\cite{Binder1981Critical}. 
In a Gaussian ordered phase, $U = 2/3$ while in a disordered phase, $U=0$. 
The Binder cumulant is especially useful for extracting $x_c$ by the crossing point method described below because it has very small finite-size corrections \cite{Shao2016Quantum}.

\begin{figure}
\begin{center}
\includegraphics[width=1.0\columnwidth]{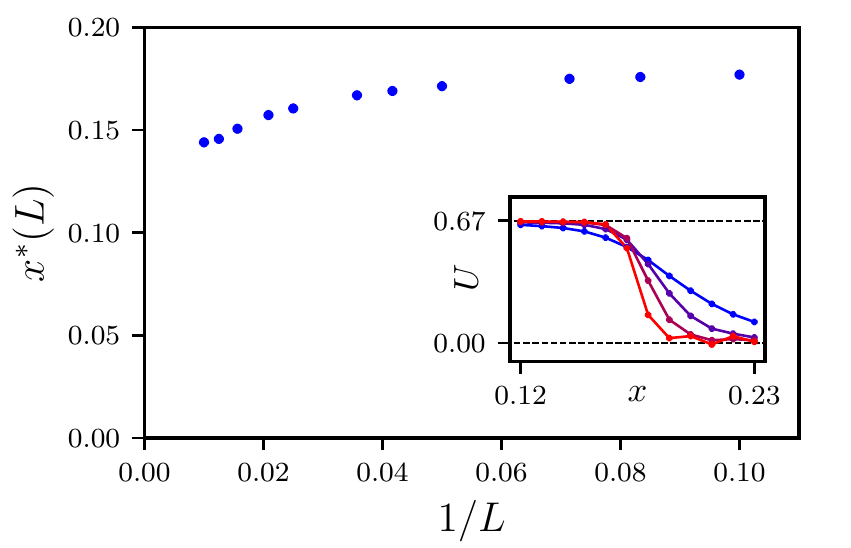}
\caption{Convergence of the Binder cumulant crossing points to the location of the sign phase transition (at $t = 10^5$). 
$x^*(L)$ is the point at which the size-$L$ and size-$2L$ cumulant curves cross, and the $1/L \rightarrow 0$ limit is the thermodynamic value $x_c$ for the transition. 
Errorbars are imperceptible on this scale. 
(inset) Representative cumulant curves for small systems (blue to red: $L = 10, 20, 40, 80$), together with the ordered and disordered limiting values of 2/3 and 0.}
\label{fig:cumulant_results}
\end{center}
\end{figure}

The inset to Fig.~\ref{fig:cumulant_results} shows representative data for the Binder cumulant $U$ computed at the longest times accessible to our simulations ($t_{\textrm{max}} = 10^5$) as a function of $x$ at several system sizes. 
At a given size $L$, $U$ crosses over from its ordered value at small $x$ to the disordered value at large $x$. 
There is significant finite size drift of the crossing points between consecutive system sizes $L$. 
The main panel of Fig.~\ref{fig:cumulant_results} shows the crossing point $x^*(L)$ for the size-$L$ and size-$2L$ curves as a function of $1/L$. 
These are obtained by fitting $U_L(x)$ as described in the Appendix.
Without further assumptions regarding the finite-size scaling of the transition, we cannot make a quantitatively accurate estimate of $x_c = \lim_{L\to\infty} x^*(L)$, but the data in Fig.~\ref{fig:cumulant_results} appear consistent with $x_c \approx 0.14$. 

To summarize, at small $x$ in 3D, the sign field orders spontaneously at long times even when initialized with a disordered configuration. 
At large $x$, on the other hand, the sign field disorders even when starting from an ordered configuration. 
As we increase the system size, the Binder cumulant of the sign field flows to the ordered limit at small $x$ and the disordered limit at large $x$, and a crossing-point analysis shows that the transition persists into the thermodynamic limit. 

These numerical results are robust but limited by finite computational resources.
Moreover, we note that sign order is in some tension with the established marginal flow of the 3D \textit{positive}-weight directed polymer to the pinned phase at arbitrarily small disorder.
We speculate that there are three possible renormalization group scenarios for sign order in 3D:
\begin{itemize}
	\item Sign order is consistent with pinned-phase fluctuations of the $\ln |A|$ field because of interference from subdominant paths. 
	%

	\item Negative amplitudes stabilize the Gaussian phase of the directed polymer in 3D and the sign-ordered phase coincides.

	\item Sign order is ultimately unstable in 3D due to the fluctuations in the strongly pinned phase. As the flow to strong pinning is only marginal, the disordering timescales are too long to be observable.
\end{itemize}
It would be very interesting to conclusively establish which of these scenarios holds and develop a theory of the associated fixed points.

\section{Applications} 
\label{sec:applications}

Our results have physical consequences for a variety of systems, which we now describe.

\subsection{Magnetoresistance of variable-range hopping}

\begin{figure}[t]
\begin{center}
\includegraphics[width=1.0\columnwidth]{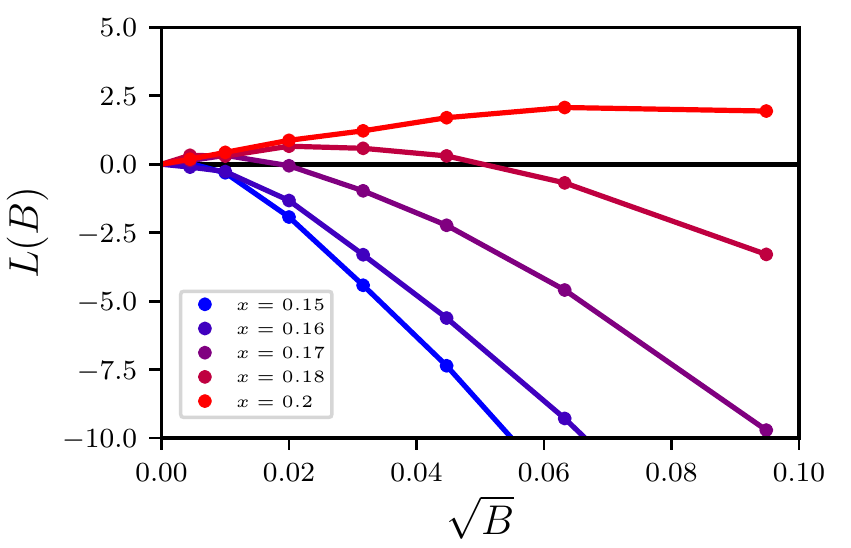}
\caption{Proxy for the 3D magnetoresistance $L(B)$ (see Eq.~\eqref{eq:magnetoresistance_measure}) as a function of the applied magnetic field $B$, for various $x$ (system size $L = 1000$ and length $t=1000$).
The solid lines are guides to the eye.}
\label{fig:magnetoresponse_3D}
\end{center}
\end{figure} 

In the variable range hopping regime of disordered semiconductors, electrons tunnel further than the typical distance between localized states~\cite{Efros1984Electronic,Mott1990Metal}.
In this case, $A$ in Eq.~\eqref{eq:path_sum} is the electron tunneling amplitude
given as a sum of partial amplitudes corresponding to different tunneling paths $\Gamma$
\cite{Shklovskii1985Effect,Nguyen1985Tunnel,Sivan1988Orbital,Medina1990Magnetic,Gangopadhyay2013Magnetoresistance}. 
In the presence of a magnetic field, each amplitude acquires a factor    
$\exp{( i\frac{\Phi_{\Gamma}}{\Phi_{0}})}$, where $\Phi_{\Gamma}=BS_{\Gamma}$ is the flux enclosed between $\Gamma$ and some fixed reference path, and $\Phi_0$ is the flux quantum (see Fig.~\ref{fig:interfering_paths}).
It has been argued that the magnetoresistance is positive in the sign-ordered phase and negative in the sign-disordered phase~\cite{Zhao1991Negative}. 
Thus, our results imply that in 2D systems at sufficiently small magnetic fields and low temperature, the magnetoresistance is always negative. 
In contrast, in 3D systems, the manetoresistance should change sign as a function of the concentration of negative scatterers $x$. 

This point is illustrated in Fig.~\ref{fig:magnetoresponse_3D}, which shows the magnetoresponse for 3D systems with the lattice of Eq.~\eqref{eq:3D_lattice_recursion}. 
We plot the quantity
\begin{equation} 
	\label{eq:magnetoresistance_measure}
	L(B) \equiv \ln{\left| \frac{A_B(0, t)}{A_0(0, t)} \right| ^2},
\end{equation}
for a fixed large value of $t$. 
$A_B(s, t)$ is the path sum in the presence of a magnetic field $B$ (see Appendix for details). 
$L(B) \sim t \Delta \xi  / \xi^2$ measures the relative change $\Delta \xi = \xi(B) - \xi(0)$ in the effective localization length which enters into the hopping conductivity.
$L(B) > 0$ indicates negative and $L(B) < 0$ indicates positive magnetoresistance (see Ref.~\cite{Zhao1991Negative} for details). 
The magnetoresistance at small $B$ indeed changes sign as a function of $x$ at $x_c$ in agreement with our estimate of the sign phase boundary.

\subsection{3D spin glass phase diagram}

\begin{figure}[t]
\begin{center}
\includegraphics[width=1.0\columnwidth]{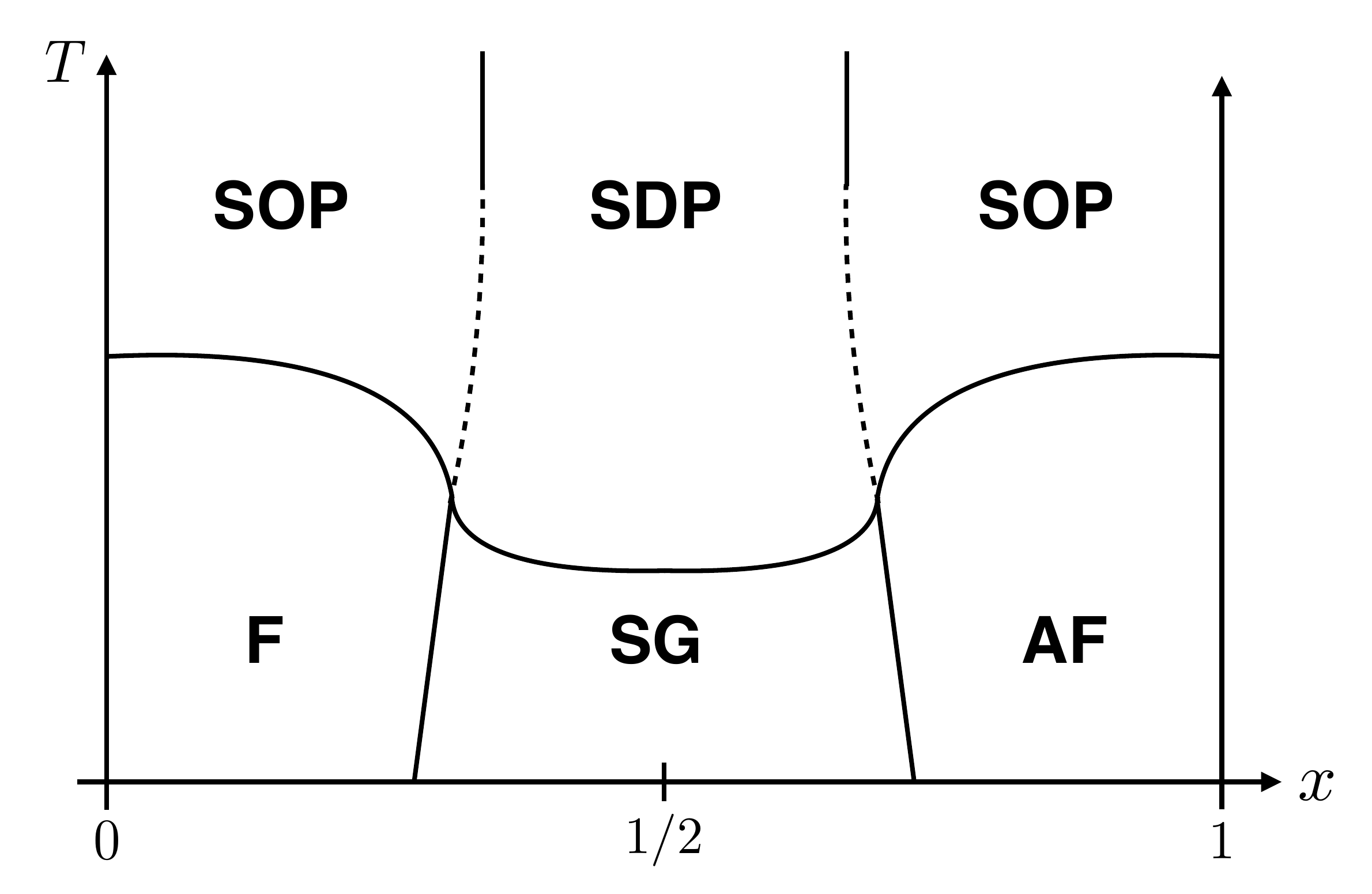}
\caption{3D spin glass phase diagram. SOP and SDP stand for the sign-ordered and sign-disordered paramagnetic phases respectively.
F, SG, and AF stand for the ferromagnetic, spin-glass, and antiferromagnetic phases respectively. Dashed lines are conjectured parts of the phase diagram.}
\label{fig:spin_glass_diagram}
\end{center}
\end{figure}

Consider a spin glass described by the Hamiltonian  
\begin{equation}\label{eq:spin_glass_hamiltonian}
	H = \sum_{ij} J_{ij} S_i S_j,
\end{equation}
where the $S_{i}$ are spins (Ising, Heisenberg, etc), and the exchange energies $J_{ij}$ are random. 
For 3D spin glasses with $\textrm{Pr}[J_{ij} > 0] = 1 - x$ and $\textrm{Pr}[J_{ij} < 0] = x$, the existence of the sign phase transition implies new features of the phase diagram. 
This is qualitatively shown in Fig.~\ref{fig:spin_glass_diagram} for the case of a bipartite (e.g. cubic) lattice. 

The three low-temperature phases (ferromagnetic, spin-glass, and antiferromagnetic) are well-established, both for Ising and Heisenberg spins \cite{LeDoussal1988Location,Hukushima2000Random,Fernandez2009Phase,Hasenbusch2007Critical,Viet2009MonteCarlo,Ceccarelli2011Ferromagnetic}. 
At high temperature $T \gg J_{ij}$, the system is paramagnetic. 
The high-temperature expansion for the spin correlation function,
\begin{equation} \label{eq:spin_correlation}
\avg{S_i S_f} = \textrm{Tr} \left[ S_i S_f \frac{\exp{(-\beta H)}}{Z} \right] ,
\end{equation}
can be expressed as a sum over interfering directed paths of the form Eq.~\eqref{eq:path_sum}. 
Thus, there is a sign phase transition in the statistical properties of $\avg{S_i S_f}$ at long distance. 
This is indicated by the vertical solid lines at high temperature in Fig.~\ref{fig:spin_glass_diagram}. 
Note that the line at $x > \frac{1}{2}$, corresponding to mainly antiferromagnetic bonds, is a transition in the N\'eel correlator $(-1)^{|i-f|} \avg{S_i S_f}$.

The sign phase transition divide phases with different symmetries in their correlators. 
Thus although we have only demonstrated the existence of the transition at high temperature, the boundary cannot terminate in the middle of the phase diagram. 
We conjecture that it meets the triple point of the thermodynamic phases (dashed lines in Fig.~\ref{fig:spin_glass_diagram}). 
In that sense, it is a continuation of the boundary separating the (anti-)ferromagnetic and spin-glass phases. 
However, all thermodynamic properties of the system are analytic across the across the sign phase transition.

It is now widely believed that the 2D equilibrium spin-glass phase does not exist at finite temperature. 
It is interesting to note that this is consistent with the non-existence of the sign-ordered phase also in 2D. 

\subsection{Composite D-wave superconductors}

\begin{figure}[t]
\begin{center}
\includegraphics[width=1.0\columnwidth]{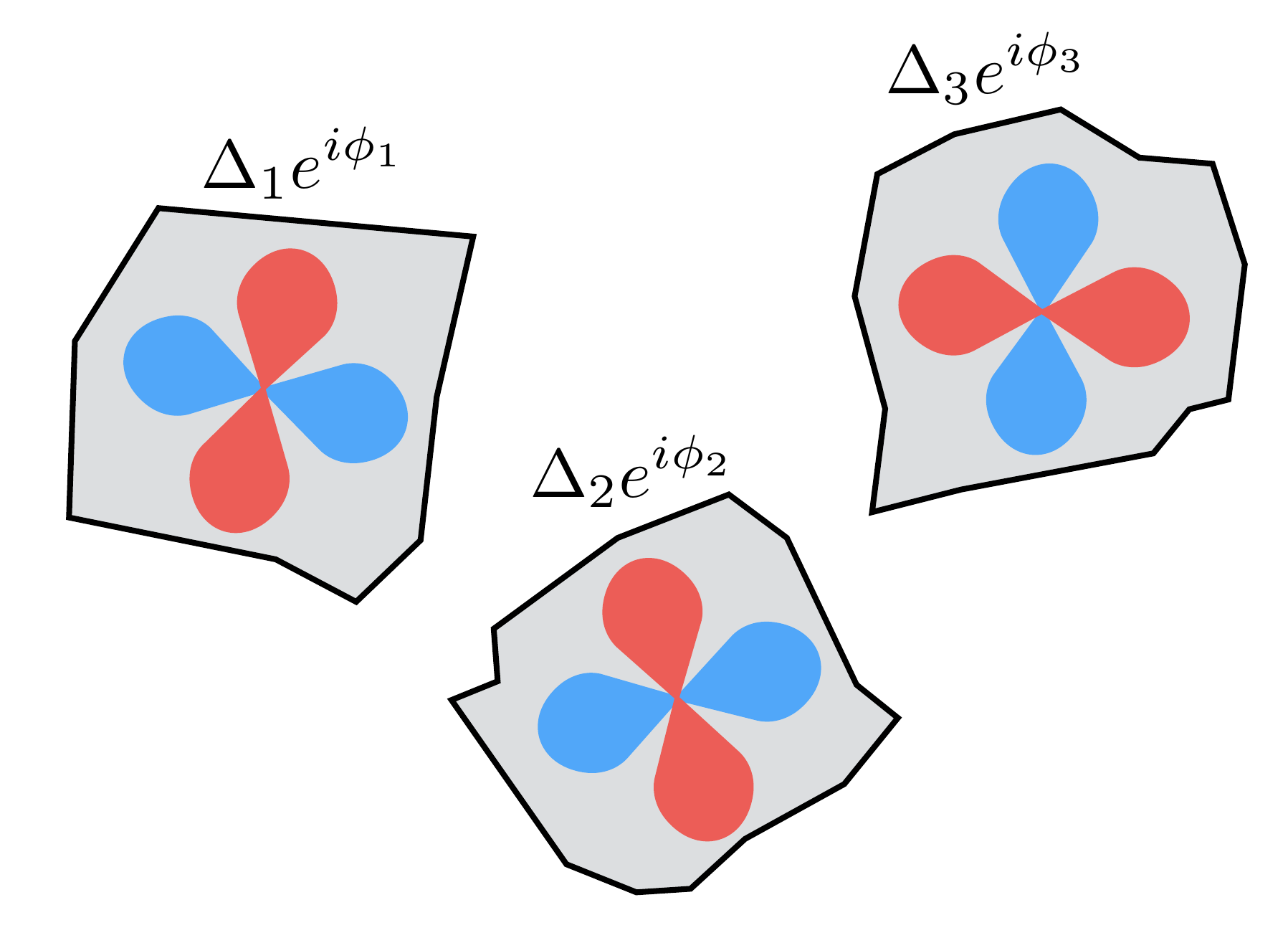}
\caption{Sketch of D-wave superconducting grains embedded in a non-superconducting medium. The lobes on each grain reflect the random alignment of the D-wave order parameter $\Delta (\textbf{k}) e^{i \phi}$.}
\label{fig:D_wave_sketch}
\end{center}
\end{figure}


The existence of sign-ordered and -disordered phases manifests in properties of 
random composite D-wave superconductors, where superconducting grains are embedded into a metallic matrix (see Fig.~\ref{fig:D_wave_sketch}). 
In the regime where the grain size is larger than the coherence length and the temperature is below the bulk $T_c$, fluctuations in the magnitude of the superconducting order parameter can be neglected.
The superconducting phases $\phi_i$ on each grain are then described by the Josephson Hamiltonian, 
\begin{equation} \label{eq:d_wave_energy_field}
H_d = \sum_{ij} J_{ij} \cos{\left( \phi_i - \phi_j + \frac{2e}{c}\int_{\textbf{r}_i}^{\textbf{r}_j} \textrm{d} \textbf{r} \cdot \textbf{A} (\textbf{r}) \right) },
\end{equation}
where $\textbf{A}(\textbf{r})$ is the vector potential.
In D-wave systems with no applied field ($\mathbf{A}(\mathbf{r}) = 0$), the effective Josephson couplings $J_{ij}$ may have random signs which depend on the separation and orientation of the embedded grains.
For further details, see the discussion in~\cite{Spivak2008Theory}. 

The high-temperature expansion of the correlation function $\chi=\avg{\exp{(i (\phi_i - \phi_j))}}$ reduces to Eq.~\eqref{eq:path_sum}, and the system can exhibit a sign phase transition as a function of the concentration of negative $J_{ij}$. 
On application of a magnetic field, this correlation function \textit{increases} in the sign-disordered phase just like the negative magnetoresistance in hopping conductivity.
This manifests as a general magnetic-field enhancement of the superconductivity.

\section{Conclusion} 
\label{sec:conclusion}

We have shown that the sign-ordered phase of the directed path sum is unstable in $d=2$ and provided strong numerical evidence that it exists in $d=3$. 
In 2D, these results have been argued previously in the regime of intermediate disorder. 
Here, we identified a large emergent length scale on which fluctuations destabilize the sign order even in the limit of weak disorder ($x \to 0$). 
The associated stretched-exponential disordering time explains the difficulty of observing disordering in previous numerical studies.

The application of our results to the 3D spin glass (Fig.~\ref{fig:spin_glass_diagram}) suggests that there are both sign-ordered and -disordered high temperature paramagnets in this canonical model. 
The nature of the phase boundaries and proposed tetracritical point require further study.
It is also an open question how best to experimentally observe the sign-ordered high temperature phase.
We plan to continue work along these lines.

The magnetic-field enhancement of superconductivity in disordered D-wave materials is an intriguing phenomenon. It would be interesting to characterize the physical regime in which this enhancement would be observable in real materials.


\section{Acknowledgements}

We would like to thank D. Huse and A. Sandvik for useful discussions and advice. CLB acknowledges the support of the NSF through a Graduate Research Fellowship, Grant No. DGE-1256082. CRL acknowledges support from the Sloan Foundation through a Sloan Research Fellowship and the NSF through Grant No. PHY-1656234.

\bibliography{SPT_Biblio}

\begin{thebibliography}{31}%
\makeatletter
\providecommand \@ifxundefined [1]{%
 \@ifx{#1\undefined}
}%
\providecommand \@ifnum [1]{%
 \ifnum #1\expandafter \@firstoftwo
 \else \expandafter \@secondoftwo
 \fi
}%
\providecommand \@ifx [1]{%
 \ifx #1\expandafter \@firstoftwo
 \else \expandafter \@secondoftwo
 \fi
}%
\providecommand \natexlab [1]{#1}%
\providecommand \enquote  [1]{``#1''}%
\providecommand \bibnamefont  [1]{#1}%
\providecommand \bibfnamefont [1]{#1}%
\providecommand \citenamefont [1]{#1}%
\providecommand \href@noop [0]{\@secondoftwo}%
\providecommand \href [0]{\begingroup \@sanitize@url \@href}%
\providecommand \@href[1]{\@@startlink{#1}\@@href}%
\providecommand \@@href[1]{\endgroup#1\@@endlink}%
\providecommand \@sanitize@url [0]{\catcode `\\12\catcode `\$12\catcode
  `\&12\catcode `\#12\catcode `\^12\catcode `\_12\catcode `\%12\relax}%
\providecommand \@@startlink[1]{}%
\providecommand \@@endlink[0]{}%
\providecommand \url  [0]{\begingroup\@sanitize@url \@url }%
\providecommand \@url [1]{\endgroup\@href {#1}{\urlprefix }}%
\providecommand \urlprefix  [0]{URL }%
\providecommand \Eprint [0]{\href }%
\providecommand \doibase [0]{http://dx.doi.org/}%
\providecommand \selectlanguage [0]{\@gobble}%
\providecommand \bibinfo  [0]{\@secondoftwo}%
\providecommand \bibfield  [0]{\@secondoftwo}%
\providecommand \translation [1]{[#1]}%
\providecommand \BibitemOpen [0]{}%
\providecommand \bibitemStop [0]{}%
\providecommand \bibitemNoStop [0]{.\EOS\space}%
\providecommand \EOS [0]{\spacefactor3000\relax}%
\providecommand \BibitemShut  [1]{\csname bibitem#1\endcsname}%
\let\auto@bib@innerbib\@empty
\bibitem [{\citenamefont {Nguyen}\ \emph {et~al.}(1985)\citenamefont {Nguyen},
  \citenamefont {Spivak},\ and\ \citenamefont {Shklovskii}}]{Nguyen1985Tunnel}%
  \BibitemOpen
  \bibfield  {author} {\bibinfo {author} {\bibfnamefont {V.~L.}\ \bibnamefont
  {Nguyen}}, \bibinfo {author} {\bibfnamefont {B.}~\bibnamefont {Spivak}}, \
  and\ \bibinfo {author} {\bibfnamefont {B.~I.}\ \bibnamefont {Shklovskii}},\
  }\href@noop {} {\bibfield  {journal} {\bibinfo  {journal} {Sov. Phys. JETP}\
  }\textbf {\bibinfo {volume} {89}},\ \bibinfo {pages} {1770 } (\bibinfo {year}
  {1985})}\BibitemShut {NoStop}%
\bibitem [{\citenamefont {Shklovskii}\ and\ \citenamefont
  {Spivak}(1991)}]{Shklovskii1991Scattering}%
  \BibitemOpen
  \bibfield  {author} {\bibinfo {author} {\bibfnamefont {B.~I.}\ \bibnamefont
  {Shklovskii}}\ and\ \bibinfo {author} {\bibfnamefont {B.~Z.}\ \bibnamefont
  {Spivak}},\ }\enquote {\bibinfo {title} {Hopping transport in solids},}\ \
  (\bibinfo  {publisher} {Elsevier Science Publishers B.V.},\ \bibinfo {year}
  {1991})\ Chap.~\bibinfo {chapter} {9}\BibitemShut {NoStop}%
\bibitem [{\citenamefont {Shapir}\ and\ \citenamefont
  {Wang}(1987)}]{Shapir1987Absence}%
  \BibitemOpen
  \bibfield  {author} {\bibinfo {author} {\bibfnamefont {Y.}~\bibnamefont
  {Shapir}}\ and\ \bibinfo {author} {\bibfnamefont {X.-R.}\ \bibnamefont
  {Wang}},\ }\href@noop {} {\bibfield  {journal} {\bibinfo  {journal} {EPL
  (Europhysics Letters)}\ }\textbf {\bibinfo {volume} {4}},\ \bibinfo {pages}
  {1165} (\bibinfo {year} {1987})}\BibitemShut {NoStop}%
\bibitem [{\citenamefont {Medina}\ \emph {et~al.}(1989)\citenamefont {Medina},
  \citenamefont {Kardar}, \citenamefont {Shapir},\ and\ \citenamefont
  {Wang}}]{Medina1989Interference}%
  \BibitemOpen
  \bibfield  {author} {\bibinfo {author} {\bibfnamefont {E.}~\bibnamefont
  {Medina}}, \bibinfo {author} {\bibfnamefont {M.}~\bibnamefont {Kardar}},
  \bibinfo {author} {\bibfnamefont {Y.}~\bibnamefont {Shapir}}, \ and\ \bibinfo
  {author} {\bibfnamefont {X.~R.}\ \bibnamefont {Wang}},\ }\href@noop {}
  {\bibfield  {journal} {\bibinfo  {journal} {Phys. Rev. Lett.}\ }\textbf
  {\bibinfo {volume} {62}},\ \bibinfo {pages} {941} (\bibinfo {year}
  {1989})}\BibitemShut {NoStop}%
\bibitem [{\citenamefont {Medina}\ and\ \citenamefont
  {Kardar}(1992)}]{Medina1992Quantum}%
  \BibitemOpen
  \bibfield  {author} {\bibinfo {author} {\bibfnamefont {E.}~\bibnamefont
  {Medina}}\ and\ \bibinfo {author} {\bibfnamefont {M.}~\bibnamefont
  {Kardar}},\ }\href@noop {} {\bibfield  {journal} {\bibinfo  {journal} {Phys.
  Rev. B}\ }\textbf {\bibinfo {volume} {46}},\ \bibinfo {pages} {9984}
  (\bibinfo {year} {1992})}\BibitemShut {NoStop}%
\bibitem [{\citenamefont {Roux}\ and\ \citenamefont
  {Coniglio}(1994)}]{Roux1994Interference}%
  \BibitemOpen
  \bibfield  {author} {\bibinfo {author} {\bibfnamefont {S.}~\bibnamefont
  {Roux}}\ and\ \bibinfo {author} {\bibfnamefont {A.}~\bibnamefont
  {Coniglio}},\ }\href@noop {} {\bibfield  {journal} {\bibinfo  {journal}
  {Journal of Physics A: Mathematical and General}\ }\textbf {\bibinfo {volume}
  {27}},\ \bibinfo {pages} {5467} (\bibinfo {year} {1994})}\BibitemShut
  {NoStop}%
\bibitem [{\citenamefont {Lien~Nguyen}\ and\ \citenamefont
  {Gamietea}(1996)}]{Nguyen1996Crossover}%
  \BibitemOpen
  \bibfield  {author} {\bibinfo {author} {\bibfnamefont {V.}~\bibnamefont
  {Lien~Nguyen}}\ and\ \bibinfo {author} {\bibfnamefont {A.~D.}\ \bibnamefont
  {Gamietea}},\ }\href@noop {} {\bibfield  {journal} {\bibinfo  {journal}
  {Phys. Rev. B}\ }\textbf {\bibinfo {volume} {53}},\ \bibinfo {pages} {7932}
  (\bibinfo {year} {1996})}\BibitemShut {NoStop}%
\bibitem [{\citenamefont {Spivak}\ \emph {et~al.}(1996)\citenamefont {Spivak},
  \citenamefont {Feng},\ and\ \citenamefont {Zeng}}]{Spivak1996Sign}%
  \BibitemOpen
  \bibfield  {author} {\bibinfo {author} {\bibfnamefont {B.}~\bibnamefont
  {Spivak}}, \bibinfo {author} {\bibfnamefont {S.}~\bibnamefont {Feng}}, \ and\
  \bibinfo {author} {\bibfnamefont {F.}~\bibnamefont {Zeng}},\ }\href@noop {}
  {\bibfield  {journal} {\bibinfo  {journal} {JETP Letters}\ }\textbf {\bibinfo
  {volume} {64}},\ \bibinfo {pages} {312} (\bibinfo {year} {1996})}\BibitemShut
  {NoStop}%
\bibitem [{\citenamefont {Aponte}\ and\ \citenamefont
  {Medina}(1998)}]{Aponte1998Directed}%
  \BibitemOpen
  \bibfield  {author} {\bibinfo {author} {\bibfnamefont {E.~G.}\ \bibnamefont
  {Aponte}}\ and\ \bibinfo {author} {\bibfnamefont {E.}~\bibnamefont
  {Medina}},\ }\href@noop {} {\bibfield  {journal} {\bibinfo  {journal} {Phys.
  Rev. E}\ }\textbf {\bibinfo {volume} {58}},\ \bibinfo {pages} {4246}
  (\bibinfo {year} {1998})}\BibitemShut {NoStop}%
\bibitem [{\citenamefont {Kim}\ and\ \citenamefont
  {Huse}(2011)}]{Kim2011Interfering}%
  \BibitemOpen
  \bibfield  {author} {\bibinfo {author} {\bibfnamefont {H.}~\bibnamefont
  {Kim}}\ and\ \bibinfo {author} {\bibfnamefont {D.~A.}\ \bibnamefont {Huse}},\
  }\href@noop {} {\bibfield  {journal} {\bibinfo  {journal} {Phys. Rev. B}\
  }\textbf {\bibinfo {volume} {83}},\ \bibinfo {pages} {052405} (\bibinfo
  {year} {2011})}\BibitemShut {NoStop}%
\bibitem [{\citenamefont {Ioffe}\ and\ \citenamefont
  {Spivak}(2013)}]{Ioffe2013Giant}%
  \BibitemOpen
  \bibfield  {author} {\bibinfo {author} {\bibfnamefont {L.}~\bibnamefont
  {Ioffe}}\ and\ \bibinfo {author} {\bibfnamefont {B.}~\bibnamefont {Spivak}},\
  }\href@noop {} {\bibfield  {journal} {\bibinfo  {journal} {Journal of
  Experimental and Theoretical Physics}\ }\textbf {\bibinfo {volume} {117}},\
  \bibinfo {pages} {551 } (\bibinfo {year} {2013})}\BibitemShut {NoStop}%
\bibitem [{\citenamefont {Shklovskii}\ and\ \citenamefont
  {Spivak}(1985)}]{Shklovskii1985Effect}%
  \BibitemOpen
  \bibfield  {author} {\bibinfo {author} {\bibfnamefont {B.~I.}\ \bibnamefont
  {Shklovskii}}\ and\ \bibinfo {author} {\bibfnamefont {B.~Z.}\ \bibnamefont
  {Spivak}},\ }\href@noop {} {\bibfield  {journal} {\bibinfo  {journal}
  {Journal of Statistical Physics}\ }\textbf {\bibinfo {volume} {38}},\
  \bibinfo {pages} {267} (\bibinfo {year} {1985})}\BibitemShut {NoStop}%
\bibitem [{\citenamefont {Sivan}\ \emph {et~al.}(1988)\citenamefont {Sivan},
  \citenamefont {Entin-Wohlman},\ and\ \citenamefont
  {Imry}}]{Sivan1988Orbital}%
  \BibitemOpen
  \bibfield  {author} {\bibinfo {author} {\bibfnamefont {U.}~\bibnamefont
  {Sivan}}, \bibinfo {author} {\bibfnamefont {O.}~\bibnamefont
  {Entin-Wohlman}}, \ and\ \bibinfo {author} {\bibfnamefont {Y.}~\bibnamefont
  {Imry}},\ }\href@noop {} {\bibfield  {journal} {\bibinfo  {journal} {Phys.
  Rev. Lett.}\ }\textbf {\bibinfo {volume} {60}},\ \bibinfo {pages} {1566}
  (\bibinfo {year} {1988})}\BibitemShut {NoStop}%
\bibitem [{\citenamefont {Medina}\ \emph {et~al.}(1990)\citenamefont {Medina},
  \citenamefont {Kardar}, \citenamefont {Shapir},\ and\ \citenamefont
  {Wang}}]{Medina1990Magnetic}%
  \BibitemOpen
  \bibfield  {author} {\bibinfo {author} {\bibfnamefont {E.}~\bibnamefont
  {Medina}}, \bibinfo {author} {\bibfnamefont {M.}~\bibnamefont {Kardar}},
  \bibinfo {author} {\bibfnamefont {Y.}~\bibnamefont {Shapir}}, \ and\ \bibinfo
  {author} {\bibfnamefont {X.~R.}\ \bibnamefont {Wang}},\ }\href@noop {}
  {\bibfield  {journal} {\bibinfo  {journal} {Phys. Rev. Lett.}\ }\textbf
  {\bibinfo {volume} {64}},\ \bibinfo {pages} {1816} (\bibinfo {year}
  {1990})}\BibitemShut {NoStop}%
\bibitem [{\citenamefont {Zhao}\ \emph {et~al.}(1991)\citenamefont {Zhao},
  \citenamefont {Spivak}, \citenamefont {Gelfand},\ and\ \citenamefont
  {Feng}}]{Zhao1991Negative}%
  \BibitemOpen
  \bibfield  {author} {\bibinfo {author} {\bibfnamefont {H.~L.}\ \bibnamefont
  {Zhao}}, \bibinfo {author} {\bibfnamefont {B.~Z.}\ \bibnamefont {Spivak}},
  \bibinfo {author} {\bibfnamefont {M.~P.}\ \bibnamefont {Gelfand}}, \ and\
  \bibinfo {author} {\bibfnamefont {S.}~\bibnamefont {Feng}},\ }\href@noop {}
  {\bibfield  {journal} {\bibinfo  {journal} {Phys. Rev. B}\ }\textbf {\bibinfo
  {volume} {44}},\ \bibinfo {pages} {10760} (\bibinfo {year}
  {1991})}\BibitemShut {NoStop}%
\bibitem [{Note1()}]{Note1}%
  \BibitemOpen
  \bibinfo {note} {In $d=1$, randomly placed single scatterers clearly disorder
  the sign field.}\BibitemShut {Stop}%
\bibitem [{\citenamefont {Fisher}\ and\ \citenamefont
  {Huse}(1991)}]{Fisher1991Directed}%
  \BibitemOpen
  \bibfield  {author} {\bibinfo {author} {\bibfnamefont {D.~S.}\ \bibnamefont
  {Fisher}}\ and\ \bibinfo {author} {\bibfnamefont {D.~A.}\ \bibnamefont
  {Huse}},\ }\href@noop {} {\bibfield  {journal} {\bibinfo  {journal} {Phys.
  Rev. B}\ }\textbf {\bibinfo {volume} {43}},\ \bibinfo {pages} {10728}
  (\bibinfo {year} {1991})}\BibitemShut {NoStop}%
\bibitem [{\citenamefont {Halpin-Healy}\ and\ \citenamefont
  {Zhang}(1995)}]{HalpinHealy1995Kinetic}%
  \BibitemOpen
  \bibfield  {author} {\bibinfo {author} {\bibfnamefont {T.}~\bibnamefont
  {Halpin-Healy}}\ and\ \bibinfo {author} {\bibfnamefont {Y.-C.}\ \bibnamefont
  {Zhang}},\ }\href@noop {} {\bibfield  {journal} {\bibinfo  {journal} {Physics
  Reports}\ }\textbf {\bibinfo {volume} {254}},\ \bibinfo {pages} {215 }
  (\bibinfo {year} {1995})}\BibitemShut {NoStop}%
\bibitem [{\citenamefont {Cugliandolo}(2010)}]{Cugliandolo2010Topics}%
  \BibitemOpen
  \bibfield  {author} {\bibinfo {author} {\bibfnamefont {L.~F.}\ \bibnamefont
  {Cugliandolo}},\ }\href@noop {} {\bibfield  {journal} {\bibinfo  {journal}
  {Physica A: Statistical Mechanics and its Applications}\ }\textbf {\bibinfo
  {volume} {389}},\ \bibinfo {pages} {4360 } (\bibinfo {year}
  {2010})}\BibitemShut {NoStop}%
\bibitem [{\citenamefont {Binder}(1981)}]{Binder1981Critical}%
  \BibitemOpen
  \bibfield  {author} {\bibinfo {author} {\bibfnamefont {K.}~\bibnamefont
  {Binder}},\ }\href@noop {} {\bibfield  {journal} {\bibinfo  {journal} {Phys.
  Rev. Lett.}\ }\textbf {\bibinfo {volume} {47}},\ \bibinfo {pages} {693}
  (\bibinfo {year} {1981})}\BibitemShut {NoStop}%
\bibitem [{\citenamefont {Shao}\ \emph {et~al.}(2016)\citenamefont {Shao},
  \citenamefont {Guo},\ and\ \citenamefont {Sandvik}}]{Shao2016Quantum}%
  \BibitemOpen
  \bibfield  {author} {\bibinfo {author} {\bibfnamefont {H.}~\bibnamefont
  {Shao}}, \bibinfo {author} {\bibfnamefont {W.}~\bibnamefont {Guo}}, \ and\
  \bibinfo {author} {\bibfnamefont {A.~W.}\ \bibnamefont {Sandvik}},\
  }\href@noop {} {\bibfield  {journal} {\bibinfo  {journal} {Science}\ }\textbf
  {\bibinfo {volume} {352}},\ \bibinfo {pages} {213} (\bibinfo {year}
  {2016})}\BibitemShut {NoStop}%
\bibitem [{\citenamefont {Efros}\ and\ \citenamefont
  {Shklovskii}(1984)}]{Efros1984Electronic}%
  \BibitemOpen
  \bibfield  {author} {\bibinfo {author} {\bibfnamefont {A.~L.}\ \bibnamefont
  {Efros}}\ and\ \bibinfo {author} {\bibfnamefont {B.~I.}\ \bibnamefont
  {Shklovskii}},\ }\href@noop {} {\emph {\bibinfo {title} {Electronic
  Properties of Doped Semiconductors}}},\ Springer Series in Solid-State
  Sciences\ (\bibinfo  {publisher} {Springer-Verlag},\ \bibinfo {year}
  {1984})\BibitemShut {NoStop}%
\bibitem [{\citenamefont {Mott}(1990)}]{Mott1990Metal}%
  \BibitemOpen
  \bibfield  {author} {\bibinfo {author} {\bibfnamefont {N.~F.}\ \bibnamefont
  {Mott}},\ }\href@noop {} {\emph {\bibinfo {title} {Metal-Insulator
  Transitions}}}\ (\bibinfo  {publisher} {Taylor \& Francis},\ \bibinfo {year}
  {1990})\BibitemShut {NoStop}%
\bibitem [{\citenamefont {Gangopadhyay}\ \emph {et~al.}(2013)\citenamefont
  {Gangopadhyay}, \citenamefont {Galitski},\ and\ \citenamefont
  {M\"uller}}]{Gangopadhyay2013Magnetoresistance}%
  \BibitemOpen
  \bibfield  {author} {\bibinfo {author} {\bibfnamefont {A.}~\bibnamefont
  {Gangopadhyay}}, \bibinfo {author} {\bibfnamefont {V.}~\bibnamefont
  {Galitski}}, \ and\ \bibinfo {author} {\bibfnamefont {M.}~\bibnamefont
  {M\"uller}},\ }\href@noop {} {\bibfield  {journal} {\bibinfo  {journal}
  {Phys. Rev. Lett.}\ }\textbf {\bibinfo {volume} {111}},\ \bibinfo {pages}
  {026801} (\bibinfo {year} {2013})}\BibitemShut {NoStop}%
\bibitem [{\citenamefont {Le~Doussal}\ and\ \citenamefont
  {Harris}(1988)}]{LeDoussal1988Location}%
  \BibitemOpen
  \bibfield  {author} {\bibinfo {author} {\bibfnamefont {P.}~\bibnamefont
  {Le~Doussal}}\ and\ \bibinfo {author} {\bibfnamefont {A.~B.}\ \bibnamefont
  {Harris}},\ }\href@noop {} {\bibfield  {journal} {\bibinfo  {journal} {Phys.
  Rev. Lett.}\ }\textbf {\bibinfo {volume} {61}},\ \bibinfo {pages} {625}
  (\bibinfo {year} {1988})}\BibitemShut {NoStop}%
\bibitem [{\citenamefont {Hukushima}(2000)}]{Hukushima2000Random}%
  \BibitemOpen
  \bibfield  {author} {\bibinfo {author} {\bibfnamefont {K.}~\bibnamefont
  {Hukushima}},\ }\href@noop {} {\bibfield  {journal} {\bibinfo  {journal}
  {Journal of the Physical Society of Japan}\ }\textbf {\bibinfo {volume}
  {69}},\ \bibinfo {pages} {631} (\bibinfo {year} {2000})}\BibitemShut
  {NoStop}%
\bibitem [{\citenamefont {Fernandez}\ \emph {et~al.}(2009)\citenamefont
  {Fernandez}, \citenamefont {Martin-Mayor}, \citenamefont {Perez-Gaviro},
  \citenamefont {Tarancon},\ and\ \citenamefont {Young}}]{Fernandez2009Phase}%
  \BibitemOpen
  \bibfield  {author} {\bibinfo {author} {\bibfnamefont {L.~A.}\ \bibnamefont
  {Fernandez}}, \bibinfo {author} {\bibfnamefont {V.}~\bibnamefont
  {Martin-Mayor}}, \bibinfo {author} {\bibfnamefont {S.}~\bibnamefont
  {Perez-Gaviro}}, \bibinfo {author} {\bibfnamefont {A.}~\bibnamefont
  {Tarancon}}, \ and\ \bibinfo {author} {\bibfnamefont {A.~P.}\ \bibnamefont
  {Young}},\ }\href@noop {} {\bibfield  {journal} {\bibinfo  {journal} {Phys.
  Rev. B}\ }\textbf {\bibinfo {volume} {80}},\ \bibinfo {pages} {024422}
  (\bibinfo {year} {2009})}\BibitemShut {NoStop}%
\bibitem [{\citenamefont {Hasenbusch}\ \emph {et~al.}(2007)\citenamefont
  {Hasenbusch}, \citenamefont {Toldin}, \citenamefont {Pelissetto},\ and\
  \citenamefont {Vicari}}]{Hasenbusch2007Critical}%
  \BibitemOpen
  \bibfield  {author} {\bibinfo {author} {\bibfnamefont {M.}~\bibnamefont
  {Hasenbusch}}, \bibinfo {author} {\bibfnamefont {F.~P.}\ \bibnamefont
  {Toldin}}, \bibinfo {author} {\bibfnamefont {A.}~\bibnamefont {Pelissetto}},
  \ and\ \bibinfo {author} {\bibfnamefont {E.}~\bibnamefont {Vicari}},\
  }\href@noop {} {\bibfield  {journal} {\bibinfo  {journal} {Phys. Rev. B}\
  }\textbf {\bibinfo {volume} {76}},\ \bibinfo {pages} {094402} (\bibinfo
  {year} {2007})}\BibitemShut {NoStop}%
\bibitem [{\citenamefont {Viet}\ and\ \citenamefont
  {Kawamura}(2009)}]{Viet2009MonteCarlo}%
  \BibitemOpen
  \bibfield  {author} {\bibinfo {author} {\bibfnamefont {D.~X.}\ \bibnamefont
  {Viet}}\ and\ \bibinfo {author} {\bibfnamefont {H.}~\bibnamefont
  {Kawamura}},\ }\href@noop {} {\bibfield  {journal} {\bibinfo  {journal}
  {Phys. Rev. B}\ }\textbf {\bibinfo {volume} {80}},\ \bibinfo {pages} {064418}
  (\bibinfo {year} {2009})}\BibitemShut {NoStop}%
\bibitem [{\citenamefont {Ceccarelli}\ \emph {et~al.}(2011)\citenamefont
  {Ceccarelli}, \citenamefont {Pelissetto},\ and\ \citenamefont
  {Vicari}}]{Ceccarelli2011Ferromagnetic}%
  \BibitemOpen
  \bibfield  {author} {\bibinfo {author} {\bibfnamefont {G.}~\bibnamefont
  {Ceccarelli}}, \bibinfo {author} {\bibfnamefont {A.}~\bibnamefont
  {Pelissetto}}, \ and\ \bibinfo {author} {\bibfnamefont {E.}~\bibnamefont
  {Vicari}},\ }\href@noop {} {\bibfield  {journal} {\bibinfo  {journal} {Phys.
  Rev. B}\ }\textbf {\bibinfo {volume} {84}},\ \bibinfo {pages} {134202}
  (\bibinfo {year} {2011})}\BibitemShut {NoStop}%
\bibitem [{\citenamefont {Spivak}\ \emph {et~al.}(2008)\citenamefont {Spivak},
  \citenamefont {Oreto},\ and\ \citenamefont {Kivelson}}]{Spivak2008Theory}%
  \BibitemOpen
  \bibfield  {author} {\bibinfo {author} {\bibfnamefont {B.}~\bibnamefont
  {Spivak}}, \bibinfo {author} {\bibfnamefont {P.}~\bibnamefont {Oreto}}, \
  and\ \bibinfo {author} {\bibfnamefont {S.~A.}\ \bibnamefont {Kivelson}},\
  }\href@noop {} {\bibfield  {journal} {\bibinfo  {journal} {Phys. Rev. B}\
  }\textbf {\bibinfo {volume} {77}},\ \bibinfo {pages} {214523} (\bibinfo
  {year} {2008})}\BibitemShut {NoStop}%
\end{thebibliography}%

\appendix

\section{Including a magnetic field in the system} 
\label{sec:appendix}

Here we focus on the 3D cubic lattice whose recursion relation is Eq.~\eqref{eq:3D_lattice_recursion}. The generalization to other geometries and dimensions is straightforward.

In general, to include a magnetic field corresponding to vector potential $\textbf{A}(\textbf{r})$, the hop from $\textbf{r}_0$ to $\textbf{r}_0 + \textbf{b}$ acquires the phase $\frac{q}{\hbar} \int_{\textbf{r}_0}^{\textbf{r}_0 + \textbf{b}} \textrm{d} \textbf{r} \cdot \textbf{A}(\textbf{r})$, where $q$ is the charge of the particle. That is,
\begin{align} \label{eq:recursion_field_modification}
& A(\textbf{r}_0 + \textbf{b}) = \alpha_{\textbf{r}_0 + \textbf{b}} A(\textbf{r}_0) + \cdots \nonumber \\
& \; \; \rightarrow A(\textbf{r}_0 + \textbf{b}) = \alpha_{\textbf{r}_0 + \textbf{b}} e^{i \frac{q}{\hbar} \int_{\textbf{r}_0}^{\textbf{r}_0 + \textbf{b}} \textrm{d} \textbf{r} \cdot \textbf{A}(\textbf{r})} A(\textbf{r}_0) + \cdots .
\end{align}
We set $q = \hbar = 1$ henceforth. 
To apply a field $\textbf{B}(s_1, s_2, t) = B \hat{\textbf{s}}_2$ to the cubic lattice, we choose vector potential $\textbf{A}(s_1, s_2, t) = - B t \, \hat{\textbf{s}}_1$ (using the orientation $\hat{\textbf{s}}_1 \times \hat{\textbf{t}} = \hat{\textbf{s}}_2$). 
The recursion relation becomes
\begin{align} \label{eq:3D_lattice_recursion_field}
A(s_1, s_2, t+1) =& \; \alpha_{(s_1, s_2, t+1)} \Big( A(s_1, s_2, t)  \nonumber \\
+ & \; e^{-i \left( t + \frac{1}{2} \right) B} A(s_1 - 1, s_2, t) \nonumber \\
+ & \; e^{i \left( t + \frac{1}{2} \right) B} A(s_1 + 1, s_2, t) \\
+ & \; A(s_1, s_2 - 1, t) + A(s_1, s_2 + 1, t) \Big) . \nonumber
\end{align}

\section{Fitting the Binder cumulant curves}

In order to accurately estimate the crossing points $x^*(L)$ of the Binder cumulant curves $U_L(x)$, we fit the data points $(x, U)$ for each $L$ to a low-order polynomial near the tentative crossing points. The ranges of $x$ over which we fit are shown in Table~\ref{tab:fit_params}. Over these intervals, second-order polynomials give acceptable fits as judged by $\chi^2$. The crossing points $x^*(L)$ shown in Fig.~\ref{fig:cumulant_results} are those of the fitted polynomials.

We estimate the uncertainty in the values $x^*(L)$ by first computing the uncertainty in the fit parameters for each curve $U_L(x)$. The data points to be fit are $(x_i, U_i)$ ($i \in \{ 1, \cdots , N\}$) with uncertainties $\sigma_i$. The fitting function is an $n$'th-order polynomial,
\begin{equation} \label{eq:fitting_poly}
U_L(x) = \sum_{k=1}^n a_k x^k,
\end{equation}
and we choose the $a_k$ to minimize
\begin{equation} \label{eq:chi_sq}
\chi^2 = \sum_{i=1}^N \left( \frac{U_L(x_i) - U_i}{\sigma_i} \right) ^2.
\end{equation}
The solution is
\begin{equation} \label{eq:chi_sq_solution}
a_k = \sum_{l=1}^n \sum_{i=1}^N ( C^{-1} ) _{kl} D_{li} U_i,
\end{equation}
where
\begin{equation} \label{eq:fit_mat_defs}
C_{kl} \equiv \sum_{i=1}^N \frac{x_i^k x_i^l}{\sigma_i^2} , \qquad D_{li} \equiv \frac{x_i^l}{\sigma_i^2}.
\end{equation}
Suppose we redo the simulation and obtain new cumulant values $U_i'$. These will deviate from $U_i$ by amounts of order $\sigma_i$. Thus the covariance matrix for the fit parameters is
\begin{equation} \label{eq:cov_mat}
\begin{aligned}
\textrm{Cov} \left[ a_k , a_l \right] =& \, \sum_{i=1}^N \sum_{j=1}^N ( C^{-1} D )_{ki} (C^{-1} D )_{lj} \textrm{Cov} \left[ U_i' , U_j' \right] \\
=& \, \sum_{i=1}^N ( C^{-1} D )_{ki} (C^{-1} D )_{li} \sigma_i^2.
\end{aligned}
\end{equation}
The underlying distribution of the fit parameters $a_k$ is unknown, but we approximate it as a Gaussian distribution with mean given by Eq.~\eqref{eq:chi_sq_solution} and covariance matrix given by Eq.~\eqref{eq:cov_mat}. Note that we have separate distributions for each system size $L$. For each pair $(L, 2L)$, we sample from the approximated distributions to obtain an ensemble of fitted curves $U_L^{(\alpha)}(x)$ \& $U_{2L}^{(\alpha)}(x)$, compute the crossing points $x^{*(\alpha)}(L)$, and take the uncertainty in $x^*(L)$ to be the standard deviation of the ensemble.

\begin{table}[t] 
\label{tab:fit_params}
\begin{tabular}{| c | c | c |}
\hline
$(L, 2L)$ & Fit range for $L$ & Fit range for $2L$ \\ \hline \hline
$(10, 20)$ & 0.174 - 0.182 & 0.168 - 0.182 \\ \hline
$(12, 24)$ & 0.172 - 0.182 & 0.171 - 0.182 \\ \hline
$(14, 28)$ & 0.170 - 0.182 & 0.169 - 0.180 \\ \hline
$(20, 40)$ & 0.168 - 0.182 & 0.167 - 0.175 \\ \hline
$(24, 48)$ & 0.166 - 0.175 & 0.164 - 0.172 \\ \hline
$(28, 56)$ & 0.162 - 0.180 & 0.162 - 0.170 \\ \hline
$(40, 80)$ & 0.158 - 0.167 & 0.155 - 0.165 \\ \hline
$(48, 96)$ & 0.153 - 0.162 & 0.152 - 0.162 \\ \hline
$(64, 128)$ & 0.145 - 0.154 & 0.145 - 0.155 \\ \hline
$(80, 160)$ & 0.140 - 0.150 & 0.140 - 0.150 \\ \hline
$(100, 200)$ & 0.138 - 0.150 & 0.139 - 0.150 \\ \hline
\end{tabular}
\caption{Fitting ranges for the Binder cumulant crossing-point analysis. First column: system sizes of the curves being fit. Second and third columns: Range of $x$ for data points being fit (spacing is $\Delta x = 0.001$).}
\end{table}


\end{document}